\documentclass[aps,twocolumn,showpacs,amsmath,amssymb,superscriptaddress,longbibliography,prl]{revtex4-1}
\usepackage[colorlinks=true,citecolor=blue,linkcolor=blue,hypertexnames=false]{hyperref}
\usepackage{amsmath}
\usepackage{dsfont}
\usepackage{hyperref}
\usepackage{textcomp}
\usepackage{tabularx}

\newcommand{\bk}{{\bf k}}

\newcommand{\bq}{{\bf q}}

\newcommand{\bE}{{\bf E}}

\newcommand{\br}{{\bf r}}

\newcommand{\bR}{{\bf R}}
\newcommand{\bK}{{\bf K}}

\newcommand{\ve}{{\varepsilon}}

\def \be{\begin{equation}}
\def \ee{\end{equation}}
\def \bea{\begin{eqnarray}}
\def \eea{\end{eqnarray}}

\begin{document}

\title{Intrinsic anomalous Hall conductivity in a nonuniform electric field}

\author{Vladyslav Kozii}
\thanks{These authors contributed equally.}
\affiliation{Department of Physics, University of California, Berkeley, CA 94720, USA}
\affiliation{Materials Sciences Division, Lawrence Berkeley National Laboratory, Berkeley, CA 94720, USA}

\author{Alexander Avdoshkin}
\thanks{These authors contributed equally.}
\affiliation{Department of Physics, University of California, Berkeley, CA 94720, USA}

\author{Shudan Zhong}
\thanks{These authors contributed equally.}
\affiliation{Department of Physics, University of California, Berkeley, CA 94720, USA}

\author{Joel E. Moore}
\affiliation{Department of Physics, University of California, Berkeley, CA 94720, USA}
\affiliation{Materials Sciences Division, Lawrence Berkeley National Laboratory, Berkeley, CA 94720, USA}

\date{\today}

\begin{abstract}
We study how the intrinsic anomalous Hall conductivity is modified in two-dimensional crystals with broken time-reversal symmetry  due to weak inhomogeneity of the applied electric field.  Focusing on a clean noninteracting two-band system without band crossings, we derive the general expression for the Hall conductivity at small finite wave vector $q$ to order $q^2$, which governs the Hall response to the second gradient of the electric field. Using the Kubo formula, we show that the answer can be expressed through the Berry curvature, Fubini-Study quantum metric, and the rank-3 symmetric tensor which is related to the quantum geometric connection and physically corresponds to the gauge-invariant part of the third cumulant of the position operator. We further compare our results with the predictions made within the semiclassical approach. By deriving the semiclassical equations of motion, we reproduce the result obtained from the Kubo formula in some limits. We also find, however, that the conventional semiclassical description in terms of the definite position and momentum of the electron is not fully consistent because of singular terms originating from the Heisenberg uncertainty principle. We thus present a clear example of a case when the semiclassical approach inherently suffers from the uncertainty principle, implying that it should be applied to systems in nonuniform fields with extra care. 
\end{abstract}

\maketitle

{\it Introduction. ---}  One of the most spectacular manifestations of quantum mechanics in solids  is the effect of band geometry and topology on transport coefficients. Examples include the anomalous Hall effect~\cite{AHEreview10}, Chern insulators~\cite{Haldane88}, topological insulators~\cite{HasanKane10,QiZhangRMP2011}, and topological semimetals~\cite{ArmitageMeleVishwanath18}.  Remarkably, some intriguing transport properties of these materials can be solely explained by the peculiarities of the band structure~\cite{XiaoChangNiureview2010}. For instance, the intrinsic anomalous Hall effect in magnetic materials can be elegantly described in terms of the Berry curvature on the Brillouin zone~\cite{ChangNiu1996,SundaramNiu1999,XiaoChangNiureview2010,Haldane04,AHEreview10}. Another interesting example is the explanation of natural optical activity at low frequencies, also known as gyrotropic magnetic effect, via the intrinsic magnetic moment of the Bloch electrons on the Fermi surface~\cite{ZhongMooreSouza2016,MaPesin2015}.

Recently, attention has been drawn to the conducting properties of metals and insulators in an inhomogeneous electric field. One set of studies focused on the momentum-dependent part of the Hall conductivity in magnetic field and its profound connection to the Hall viscosity in Galilean invariant and lattice systems~\cite{HoyosSon2012,BradlynGoldsteinRead12,Scaffidietal2017,Gromov2017,HolderQueirozStern2019,HarperJacksonRoy2018,ChenJiangBiswas2018}. Moreover, it was shown that the Hall viscosity determines the size-dependent  part of the Hall resistance and can be used as an indication of the hydrodynamic flow of the electron fluid~\cite{Scaffidietal2017,Gromov2017,HolderQueirozStern2019}, which was recently probed experimentally~\cite{Berdyuginetal2019}. Other works studied the modifications of the semiclassical equations of motion in inhomogeneous field due to the nonzero Fubini-Study quantum metric and the manifestation of these modifications in transport and optical measurements~\cite{LapaHughes2019,GaoXiao2019}.

In this work, we study the intrinsic contribution to the anomalous Hall current in  a nonuniform electric field. Instead of  an external magnetic field, we assume that time-reversal symmetry in a crystal is broken by, e.g., magnetic order, leading to nonzero Berry curvature. The correction due to the electric field inhomogeneity is captured by the momentum dependence of the Hall conductivity, which at small momenta $q$ can be expanded as (choosing vector $\bq$ to be along the $x$-axis)

\be  
\sigma_{AH}(q) = \sigma_{AH}^{(0)} + q^2 \sigma_{AH}^{(2)} + \ldots \label{Eq:sigmaAH}
\ee
Here we explicitly defined the antisymmetric part of the conductivity tensor as $\sigma_{AH}(\bq) \equiv [\sigma_{xy}(\bq) - \sigma_{yx}(\bq)]/2$, and the subscript ``$AH$'' stands for ``anomalous Hall''~\footnote{The $q$-linear correction to the conductivity tensor has been studied in detail in Refs.~\cite{LapaHughes2019,GaoXiao2019}. It is symmetric, so it does not appear in Eq.~\eqref{Eq:sigmaAH}, and requires breaking of both time-reversal and inversion symmetries.}. 
%We focus on two-dimensional systems as in previous work on quantum Hall systems, so the order-$q_z$ term in the expansion~\cite{ZhongMooreSouza2016,MaPesin2015} is absent.
We calculate $\sigma_{AH}^{(2)}$ for a generic clean two-dimensional two-band system (without any external magnetic field) and show that it is expressed through three gauge-invariant objects defined on the Brillouin zone: Berry curvature, quantum metric, and the fully symmetric rank-3 tensor  defined through the symplectic connection~\cite{Fedosov:1996fu}.  The latter object determines the gauge-invariant part of the third cumulant of the position operator (analogous to how quantum metric determines the second cumulant of the position operator) and was recently shown to enter the answer for the shift photocurrent in Weyl semimetals~\cite{Patankar_2018,AhnNagaosa2020}. We assume that the two bands are separated by a finite energy gap everywhere in the Brillouin zone.  We use the Kubo formula to obtain the most general microscopic answer.

We further compare our result with the answer obtained within the semiclassical approach. To do that, we derive the semiclassical equations of motion in a nonuniform electric field up to the second order in the field gradients. We find that semiclassics reproduces the terms dominating $\sigma_{AH}^{(2)}$ in the insulating regime in the limit when the two bands are well separated. However, we also show that the Heisenberg uncertainty principle does not allow for a reliable semiclassical description in terms of the wave packet dynamics when dealing with the second gradients of the electric field. In particular, we find that the equations of motion for the electron wave packet contain  some terms which formally become divergent in the case when the wave packet is narrow in momentum space (i.e., corresponds to the state with a well-defined momentum). The origin of these terms is clear: the semiclassical Boltzmann approach relies on the assumption that the particles are described by the well-defined momentum and coordinate, which is inherently incompatible with the basics of quantum mechanics. This contradiction does not cause serious problems when dealing with the uniform part of the electric field or with its first gradient, and we also show that the semiclassics well agrees with the Kubo formula in some limits once these wave packet-dependent terms are discarded. However, our work presents a clear example of how the uncertainty principle dramatically reveals itself in the conventional semiclassical description of the systems in inhomogeneous fields, imposing the natural limits of its applicability.

%This contradiction, however, does not cause any problems when dealing with the uniform part of the electric field or with its first gradient, and we also show that the semiclassics well agrees with the Kubo formula in some limits once these wavepacket-dependent terms are discarded. 

{\it Kubo formula result. ---} The  most straightforward way to calculate the intrinsic Hall conductivity from the microscopic band structure is to use the Kubo formula.  For simplicity, we consider generic two-band Hamiltonian $\hat H_0(\bk)$ defined in the two-dimensional quasimomentum space, which is diagonal in the basis of Bloch wavefunctions $|u_{V,C}(\bk)\rangle$ with the spectrum $\ve_{V,C}(\bk)$, $ \hat H_0(\bk)|u_{V,C}(\bk)\rangle = \ve_{V,C}(\bk)|u_{V,C}(\bk)\rangle$. Indices ``$V$'' and ``$C$'' stand for the ``valence'' and ``conduction'' band, respectively.  

The Hall conductivity is given by the antisymmetric part of the conductivity tensor $\sigma_{\alpha \beta}(i\omega_n, \bq)$ which is related to the current-current correlation function, $K_{\alpha \beta}(i\omega_n, \bq) = \langle \hat j^\alpha(i\omega_n, \bq) \hat j^\beta(-i\omega_n, -\bq)  \rangle$, as $\sigma_{\alpha \beta}(i\omega_n, \bq) =  -K_{\alpha \beta}(i\omega_n, \bq)/\omega_n$~\cite{Colemanbook}. Strictly speaking, the correlator $K_{\alpha \beta}$ only describes the paramagnetic contribution to conductivity and, in principle, the diamagnetic term should also be included. The latter, however, does not contribute to the (antisymmetric) Hall component of conductivity and will be omitted hereafter.

We find that the simplest and most physically intuitive result is obtained in the case when  the chemical potential lies within the band gap, i.e., when the system is in the insulating state. The uniform part of the anomalous Hall conductivity is then quantized and given by $\sigma_{AH}^{(0)} = -\frac{e^2}{\hbar} \frac1S\sum_{\bk} \Omega_{xy}(\bk) = \frac{e^2}{h}C$, where $S$ is the total area of the system, integer $C$ is the Chern number, and $\Omega_{ij}(\bk) =- 2 \, \text{Im}\langle \partial_{k_i} u_V|\partial_{k_j} u_V\rangle$ is the Berry curvature of the valence band~\cite{AHEreview10}. As for the $q^2$ component of the Hall conductivity, we find that in the static limit $\omega\to 0$ it equals to %{\vk (Check overall minus and the factor $1/S$, where $S$ is the total area.)}

\begin{widetext}
\be \label{Eq:sigma02}
\sigma_{AH}^{(2)} = \frac{e^2}{2\hbar} \frac1S \sum_\bk g_{xx} \Omega_{xy}  -
  \frac{\hbar}{\ve_C - \ve_V}
\left[   \frac{v_{Cx} - v_{Vx}}3 \frac{\partial \Omega_{xy}}{\partial k_x} + \frac{v_{Cx} - v_{Vx}}2 T_{xxy} - \frac{v_{Cy} - v_{Vy}}2 T_{xxx}  \right] - \frac{2 \hbar^2 v_{Cx}v_{Vx}}{(\ve_C - \ve_V)^2}\Omega_{xy},
\ee
\end{widetext}
where the summation is over the states in the completely filled valence band and we fix our coordinate system such that  $\bq$ is along the $x$-axis. We also suppressed the indices $\bk$ in the above expression for brevity. Band velocities $v_{V(C)}$ and the quantum metric tensor of the valence band $g_{ij}$ are defined as $v_{V(C) i}(\bk) =\partial_{k_i}\ve_{V(C)}(\bk)/\hbar$ and $g_{ij}(\bk) = \text{Re}\left[\langle \partial_{k_i} u_V|\partial_{k_j} u_V\rangle - \langle \partial_{k_i} u_V|u_V\rangle \langle u_V | \partial_{k_j} u_V\rangle  \right]$, respectively.  The answer for $\sigma_{AH}^{(2)}$ in Eq.~\eqref{Eq:sigma02} also contains the components of a fully symmetric tensor 

\be
T_{ijl} =   \frac13 \, \text{Im}(c_{ijl} + c_{jli} + c_{lij}), 
\ee
where
\be
c_{ijl} = \langle u_V|(\partial_{k_i} \partial_{k_j}P_C)(\partial_{k_l} P_C)|u_V  \rangle 
\ee
is the quantum geometric connection of the valence band~\cite{AhnNagaosa2020} and $P_C  = |u_C\rangle \langle u_C| = 1-|u_V\rangle \langle u_V|$ is the projector onto the conduction band. The real part of the tensor $c_{ijl}$ is just the Christoffel symbols  of the quantum metric, while the imaginary part was identified as the symplectic  Christoffel symbols in Ref.~\cite{AhnNagaosa2020}. All the geometric quantities that determine the answer in Eq.~\eqref{Eq:sigma02}, $\Omega_{ij}$, $g_{ij}$, and $T_{ijl}$, are invariant under the gauge transformation $|u(\bk)\rangle \to e^{i\phi_\bk}|u(\bk) \rangle$.

Equation~\eqref{Eq:sigma02} is one of the main results of the present work. We see that in the case when the bandwidth is much smaller than the band gap, i.e., when bands are nearly flat, $\sigma_{AH}^{(2)}$ is mainly determined by the first term, involving the product of the Berry phase $\Omega_{xy}$ and quantum metric $g_{xx}$.  That this term indeed dominates $\sigma_{AH}$ in this limit has been verified numerically for a Haldane model with flattened bands~\cite{Zhong_thesis}.  This result can be qualitatively understood as follows. The size of the maximally localized Wannier orbital is known to be given by the quantum metric tensor $g_{ij}$~\cite{MarzariVanderbilt97,PrunedaSouza09}. The effective electric field averaged over the size of the  wave packet that the particle experiences in a slowly varying electric field can be estimated as $\bE(0)+[\partial^2 \bE(0)/\partial x^2]g_{xx}$. The correction to the anomalous velocity then equals $\delta v_j \sim \delta E_i \, \Omega_{ij} \sim  (\partial^2 E_i/\partial x^2)g_{xx} \Omega_{ij},$ which in Fourier space gives exactly $\sigma_{AH}^{(2)} \propto q^2 g_{xx} \Omega_{xy}$.

Since the tensor $T_{ijl}$ is not well known in the condensed matter context, we briefly comment on its significance. Defined as the fully symmetric part of the symplectic Christoffel symbols, $T_{i j l}$ encapsulates  certain geometric information about the band structure  similar to the Berry curvature  and quantum metric~\cite{SM}. Physically, it determines the gauge-invariant part of the third cumulant (skewness) of the position operator averaged over the electron configuration. This observation reconciles the results of Refs.~\cite{Patankar_2018} and~\cite{AhnNagaosa2020}   which computed the circular shift photocurrent in topological semimetals and expressed the answers in terms of the third cumulant and the symplectic Christoffel symbols, correspondingly. We also note that the real part of the quantum geometric connection, the Christoffel symbols, was shown to determine the linear shift photocurrent~\cite{AhnNagaosa2020}.

As an example, we consider the case of a massive Dirac fermion described by the Hamiltonian $\hat H_0(\bk) = \hbar v_F(k_x \sigma_x + k_y \sigma_y) + \Delta \sigma_z$, where $\sigma_i$ are  the Pauli matrices. We find that its Hall conductivity is given by 

\be
\sigma_{AH}^{\text{Dirac}}(q) \approx -\frac{e^2}{2h}\left(1-\frac{\hbar^2 v_F^2 q^2}{12 \Delta^2}\right).
\ee
An extra prefactor $1/2$ appears due to the fact that the realistic band structure (e.g., a Haldane model) always contains an even number of Dirac points. We further emphasize that this result as well as Eq.~\eqref{Eq:sigma02} are obtained for a system with the chemical potential inside the band gap (an insulator). In principle, one can generalize the answer for a metallic system with the chemical potential residing in a partially filled band. In this case, the final result also contains the contributions from the vicinity of the Fermi surface that are rather complicated and require extra care. In particular, these contributions are very sensitive to the order in which frequency $\omega$ and wave vector $q$ are taken to zero and demonstrate singular dependence  on the electron's density in the clean limit. These and related questions are discussed in more detail in the Supplemental Materials~\cite{SM}.

It is  known that, for the Galilean invariant quantum Hall states, $\sigma_{AH}^{(2)}$ is determined by the Hall viscosity at large magnetic fields~\cite{HoyosSon2012,BradlynGoldsteinRead12,Scaffidietal2017}. However, the direct comparison of our result for the anomalous Hall conductivity, Eq.~\eqref{Eq:sigma02}, and Hall viscosity for the lattice systems found in Ref.~\cite{RaoBradlyn2020} does not reveal any obvious connection between these two quantities. This is not surprising since a generic crystal does not possess Galilean invariance.

{\it Semiclassical description. ---} To get more intuition about the answer obtained within the Kubo formula, we now apply the semiclassical approach to the same problem. While we show that this approach is useful for obtaining certain insight into the origin of the most relevant terms in some limits, it still has a number of limitations which do not allow for an accurate quantitative description. The most restrictive limitation is imposed by the uncertainty principle. This principle  forbids a quantum particle to have a definite position and momentum simultaneously, which, in turn, is the key assumption of the semiclassical Boltzmann formalism.

We consider an electron moving in a periodic potential of a lattice with the Hamiltonian $\hat H_0$ in an inhomogeneous static electric field  $\bE(\br) = -\nabla \phi(\br)$, such that the full Hamiltonian is given by 

\be  
\hat H = \hat H_0 - e\varphi(\hat \br).
\ee 
Hamiltonian $\hat H_0(\bk)$ used in the Kubo-formula derivation is the second-quantized version of $\hat H_0$, written in momentum space. In our further derivation, we closely follow the approach of Ref.~\cite{LapaHughes2019}. In particular, we assume that the periodic part of the Hamiltonian (without the electrostatic potential), $\hat H_0$,  is diagonal in the basis of the Bloch wavefunctions $|\psi_\bk\rangle = e^{i \bk \cdot \hat\br} |u(\bk)\rangle$, $\hat H_0 |\psi_\bk\rangle  = \ve_\bk |\psi_\bk\rangle$, where $\hbar \bk$ is quasimomentum and the function $u_\bk(\br) \equiv \langle \br |u(\bk)\rangle$ has the periodicity of the crystal in real space, with the normalization condition $\langle \psi_{\bk'} | \psi_\bk\rangle = \delta(\bk - \bk')$. 

The goal now is to derive the corrections to the semiclassical equations of motion due to the finite gradients of the electric field. More specifically, we are interested in the second gradient, which is equivalent to the $q^2$ term in the Hall conductivity calculated above. To obtain the correction, we consider the dynamics of the wave packet constructed of the states within the same band and defined as $|\Psi(t)\rangle = \int d \bk \, a(\bk,t)|\psi_\bk\rangle$. Within this single-band approximation, the Schr\"{o}dinger equation $i \hbar (\partial/\partial t)|\Psi(t)\rangle = \hat H |\Psi(t)\rangle$ determines the dynamics of $a(\bk, t)$ as

\be 
i\hbar \frac{\partial a(\bk, t)}{\partial t} = \ve_\bk a(\bk,t) -e\int d \bk' \, a(\bk',t) \langle \psi_\bk|\varphi(\hat \br)|\psi_{\bk'}\rangle,
\ee 
with the normalization condition $\int d\bk |a(\bk,t)|^2 = 1$~\cite{LapaHughes2019}.

To take into account weak inhomogeneity of the electric field, we expand the electrostatic potential near $\br = 0$ as
$
\varphi (\br) = -E^\mu r_\mu - \frac12 E^{\mu \nu} r_\mu r_\nu - \frac16 E^{\mu \nu \xi} r_\mu r_\nu r_\xi - \ldots,
$
where $E^{\mu \nu \xi}$, $E^{\mu \nu}$, and $E^{\mu}$ are fully symmetric tensors that do not depend on $\br$, and the summation over the repeated indices is implied. The electric field near $\br = 0$ is then given by $E^{\mu}(\br) \approx E^\mu + E^{\mu \nu} r_\nu + \frac12 E^{\mu \nu \xi} r_\nu r_\xi$.  The correction to the electron's velocity proportional to $E^{\mu \nu \xi}$ determines  the $q^2$ term in the Hall conductivity, $\sigma_{AH}^{(2)}$.

To derive the semiclassical expression for the wave packet velocity, we define its position $R_{\alpha}(t)$ and momentum $K_\alpha (t)$  as

\begin{align}
 R_\alpha(t) &\equiv \langle \Psi(t)|\hat r_\alpha| \Psi(t)   \rangle, \nonumber \\  K_\alpha(t) &\equiv \langle \Psi(t)|\hat k_\alpha|\Psi(t)\rangle,
\end{align}
where $\hbar \hat k_\alpha$ is the quasimomentum operator satisfying $\hat k_\alpha |\psi_\bk\rangle = k_\alpha |\psi_\bk \rangle$.
Parametrizing function $a(\bk,t)$ as $a(\bk,t) = |a(\bk,t)|e^{-i \gamma(\bk,t)}$, one easily finds that 

\begin{align}
 R_\alpha(t) &= \int d\bk \, \tilde R_\alpha(\bk, t) |a(\bk, t)|^2, \nonumber \\  K_\alpha(t) &= \int d\bk \,  k_\alpha |a(\bk, t)|^2,
\end{align}
with 

\be
\tilde R_\alpha(\bk, t) \equiv \frac{\partial \gamma(\bk,t)}{\partial k_\alpha} + A_\alpha(\bk),
\ee
and $A_\alpha(\bk) = i \langle u(\bk) | \partial_{k_\alpha} u(\bk)  \rangle$ is the Berry connection. If the wave packet is strongly peaked at momentum $\bK$, $|a(\bk, t)|^2 \approx \delta(\bk - \bK)$, semiclassical coordinate of the wave packet becomes simply $R_\alpha \approx \tilde R_\alpha (\bK)$.

Thus far, our semiclassical analysis was similar to that of Ref.~\cite{LapaHughes2019}. In what follows, however, we are mostly interested in the second-order gradient correction to the semiclassical equations of motion which, to the best of our knowledge, has never been studied before. Assuming that the wavepacked is narrowly peaked in the momentum space, we find up to the order $\partial^2 \bE(\bR)/\partial R_\mu \partial R_\nu$ (equivalently, up to the order $E^{\mu \nu \xi}$):

\begin{align} \label{Eq:EOM}
\dot K^\alpha(t) &= -\frac{e}{\hbar}E^\alpha(\bR) - \frac{e}{2\hbar}E^{\alpha \mu \nu}\left[g_{\mu \nu}(\bK) +f_{\mu \nu}\{|a(\bk,t)|\}    \right], \nonumber \\ \dot R_\alpha(t) &= \frac1{\hbar}\frac{\partial \ve_\bk}{\partial k_\alpha} - \frac{e}{\hbar} E^{\mu}(\bR) \Omega_{\mu \alpha} + \frac{e}{2\hbar}\frac{\partial E^{\mu}(\bR)}{\partial R_\nu} \frac{\partial g_{\mu \nu}}{\partial K_\alpha} - \nonumber \\ &-\frac{e}{6\hbar}E^{\mu \nu \xi}\left(3 g_{\mu \nu} \Omega_{\xi \alpha} - \frac{\partial T_{\mu \nu \xi}}{\partial K_\alpha} -\frac{\partial^2 \Omega_{\xi \alpha}}{\partial K_\mu \partial K_\nu} \right) - \nonumber \\ & - \frac{e}{2\hbar}E^{\mu \nu \xi}\tilde f_{\mu \nu \xi \alpha}\left\{|a(\bk,t)|  \right\},
\end{align}
with functionals $f_{\mu \nu}$ and $\tilde f_{\mu \nu \xi \alpha}$ given by

\begin{align} \label{Eq:f}
&f_{\mu \nu}\{ |a(\bk,t)| \} \equiv \int d\bk \, \frac{\partial|a(\bk,t)|}{\partial k_\mu} \frac{\partial|a(\bk,t)|}{\partial k_\nu}, \\ &\tilde f_{\mu \nu \xi \alpha}\{ |a(\bk,t)| \} \equiv \int d\bk \, \frac{\partial|a(\bk,t)|}{\partial k_\mu} \frac{\partial|a(\bk,t)|}{\partial k_\nu}\Omega_{\xi \alpha}(\bk). \nonumber
\end{align}

Equations~\eqref{Eq:EOM} represent the second main result of the present work. The derivation is straightforward but tedious, so we delegate it to the Supplemental Materials~\cite{SM}. The first gradient correction, $E^{\mu \nu} \partial g_{\mu \nu}/\partial K_\alpha$, has been obtained and discussed in Refs.~\cite{LapaHughes2019,GaoXiao2019}, and our answer agrees with it. The second-order gradient term containing $E^{\mu \nu \xi}$ is a new result that deserves further discussion.

Within the kinetic equation approach, current density equals $j_\alpha = -(e/S)\sum_{\bK} \dot R_\alpha f_\bK$, where $f_\bK$ is the distribution function. To the leading order, $f_\bK$ is given by the Fermi-Dirac distribution, and the current is given by $-e\dot R_\alpha$ summed over the filled states, thus allowing for a simple interpretation of all the terms in Eq.~\eqref{Eq:EOM}.

One can easily show that the term with $g_{\mu \nu} \Omega_{\xi \alpha}$ exactly reproduces the first term in Eq.~\eqref{Eq:sigma02}. The remaining terms in Eq.~\eqref{Eq:sigma02} contain first or second powers of the band gap $(\ve_C - \ve_V)$ in the denominator and could, in principle, be perturbatively captured by the semiclassical approach~\cite{GaoYangNiu14,GaoXiao2019}; we, however, do not present such analysis in this work. The other two terms in Eq.~\eqref{Eq:EOM} contain full derivatives $\partial T_{\mu \nu \xi}/\partial K_\alpha$ and $\partial^2\Omega_{\xi \alpha}/\partial K_\mu \partial K_\nu$, consequently, their contribution to the current vanishes in the case of a completely filled band, so they do not appear in Eq.~\eqref{Eq:sigma02}~\footnote{These terms, however, do contribute to the Hall conductivity in the metallic regime, though the full answer contains some extra terms which are not captured by the semiclassical approach, see~\cite{SM}  for details.}.

Finally, there are terms $f_{\mu\nu}$ and $\tilde f_{\mu \nu \xi \alpha}$ in Eq.~\eqref{Eq:EOM} which  pose the main problem for the semiclassical description of the wave packet dynamics to the second order in the electric field gradients. These terms are given by Eq.~\eqref{Eq:f} and are very non-universal in a sense that they strongly depend on the shape of the wave packet, i.e., function $a(\bk,t)$. To estimate the magnitude of these terms, we may assume that $a(\bk,t)$ has form of the Gaussian distribution with the width $\Delta k$. It is clear then that $f_{\mu \nu} \propto 1/(\Delta k)^2$  and $\tilde f_{\mu \nu \xi \alpha} \propto \Omega_{\xi \alpha}/(\Delta k)^2$, thus diverging as $\Delta k \to 0$, which corresponds to the limit of well-defined quasiparticles in momentum space. These terms  originate from the correlators $\langle \Psi(t)|\hat r_\mu \hat r_\nu|\Psi(t)\rangle$ (and higher moments) and  clearly represent the Heisenberg uncertainty  principle, which implies that the wavefunctions strongly localized in  momentum space experience large variation with the position. While this fundamental principle is not an obstacle for the quaiclassical description at the zeroth and first order in field gradients, it clearly manifests itself at the second order. We see, however, that once the terms $f_{\mu\nu}$ and $\tilde f_{\mu \nu \xi \alpha}$ are neglected, our semiclassical answer well agrees with the Kubo formula calculation for an insulating case in the limit when the band separation is much larger than the bandwidth.

It is also instructive to demonstrate an alternative derivation of Eq.~\eqref{Eq:EOM}, which is less rigorous but more physically intuitive. The first equation is simply the Newton's law stating that the rate of the momentum change equals the external force: $\hbar \dot \bK = -e\langle \Psi(t)| \bE(\hat \br)|\Psi(t)\rangle$. To derive the second equation, we introduce the effective quasiparticle energy $\varepsilon_{eff}(\bR, \bK) = \langle \Psi(t)| \hat H_0 -e \varphi(\hat \br)|\Psi(t)\rangle$, where $\hbar \bK$ is the momentum of the wave packet. The equation for the effective velocity then reads as $\hbar \dot R_\alpha \approx (\partial \ve_{eff}/\partial K_\alpha) - \hbar \Omega_{\alpha \mu} \dot K_\mu$, while the Newton's law can be rewritten as $\hbar \dot \bK \approx - \partial \ve_{eff}/\partial \bR$~\cite{LapaHughes2019}. It is straightforward to check that the resulting equations are equivalent to Eq.~\eqref{Eq:EOM}. The only subtle difference originates from the singular terms analogous to Eq.~\eqref{Eq:f}, which we discuss in more detail in the Supplemental Materials~\cite{SM}.

This approach has the further advantage of elucidating the physical meaning and origin of different terms. For example, the singular terms $f_{\mu \nu}$ and $\tilde f_{\mu \nu \xi \alpha}$ originate from the correlator $\langle \Psi| \hat r_\mu \hat r_\nu |\Psi\rangle$, which determines the real-space width of the state and appears in the expression for $\hbar \dot \bK$.
While these terms are singular for the wave packets narrowly peaked in momentum space, they vanish in case of maximally localized Wannier functions, $|a(\bk)| = \text{const}$. In the latter case, the correlator can be roughly estimated by the averaged quantum metric $g_{\mu \nu}$~\cite{MarzariVanderbilt97,PrunedaSouza09}. In fact, there is a well-established procedure for how to define the width in such a way that the corresponding cumulant averaged over the filled band does not suffer from any divergencies and is given exactly by the quantum metric averaged over the same filled band~\cite{SouzaWilkensMartin2000,Sgiarovelloetal2001}. If for some reason further development of the semiclassical approach is necessary, it seems likely that this approach would allow for a formulation which is free of any singularities and completely agrees with the Kubo formula results found in this work.
%We, however, do not study this question in detail here and leave it for future publications.
Finally, when calculating $\ve_{eff}$, we notice that the tensor $T_{\mu\nu\xi}$ determines the gauge-invariant part of the third cumulant of the position operator $\hat \br$,  $T_{\mu\nu\xi} \approx \langle\Psi|\delta \hat r_\mu \delta \hat r_\nu \delta \hat r_\xi |\Psi  \rangle_{\text{g.-i.}}$, where $\delta \hat r_\mu \equiv \hat r_\mu - \langle \hat r_\mu \rangle$~\cite{SM}.

{\it Conclusions. ---} We have calculated the $q^2$ contribution to the intrinsic anomalous Hall conductivity in the inhomogeneous electric field in clean crystals without time-reversal symmetry. To do that, we have applied the Kubo formula to a generic two-band model and then compared the results with the predictions obtained from the semiclassical approach. We showed that the two approaches agree with each other in some limits once the uncertainty principle limitations of the semiclassics are neglected. We expect that this new contribution can be directly probed by the high-precision optical measurement, thus providing valuable information about the geometry of the band structure. As a next step, it would be interesting to relate the newly found $q^2$ correction to the Hall current to possible experiments revealing the hydrodynamics of electrons in solids~\cite{Varnavidesetal2020}.  In the Supplemental Materials~\cite{SM} we show that, under certain conditions, $\sigma_{AH}^{(2)}$  determines the finite size correction to the Hall resistance in the hydrodynamic regime in  crystals with broken time-reversal symmetry, analogous to how the Hall viscosity $\eta_{xy}$ does it in  the narrow channel or Corbino geometry experiments in the Galilean-invariant systems in a nonquantized external magnetic field~\cite{Scaffidietal2017,Gromov2017,HolderQueirozStern2019}. We leave the comprehensive study of these and related questions for future work.

{\it Acknowledgements. ---} We thank Vir Bulchandani, Daniel Parker, Fedor Popov, and Jonathan Ruhman for stimulating discussions. This work was supported by the Quantum Materials program at LBNL, funded by the
US Department of Energy under Contract No. DE-AC02-
05CH11231 (V.~K. and J.~E.~M.), the National Science
Foundation under Grant No. DMR-1918065 (A.~A.), a Kavli ENSI fellowship at UC Berkeley (A.~A.), and a Simons Investigatorship (J.~E.~M.).

\bibliography{bibliography}

\newpage

\begin{widetext}

\begin{center}
\textbf{\large Supplemental Materials for ``Intrinsic anomalous Hall conductivity in a nonuniform electric field''} 
\end{center}
%%%%%%%%%% Merge with supplemental materials %%%%%%%%%%
%%%%%%%%%% Prefix a "S" to all equations, figures, tables and reset the counter %%%%%%%%%%
\setcounter{equation}{0}
\setcounter{figure}{0}
\setcounter{table}{0}

\makeatletter
\renewcommand{\theequation}{S\arabic{equation}}
\renewcommand{\thefigure}{S\arabic{figure}}
\renewcommand{\thetable}{S\Roman{table}}

This Supplemental Material consists of two sections. In the first Section, we provide the details of the intrinsic Hall conductivity calculation using the Kubo formula  and show how the gradient correction can be probed experimentally in the hydrodynamic flow through a narrow channel. The second Section is dedicated to the semiclassical description in terms of the wave packet dynamics and comparison between the two approaches. 

%{\color{red} Finally, in the third Section, we show that, under certain conditions, $\sigma_{AH}^{(2)}$ determines the width-dependent correction of the Hall resistance in the hydrodynamic flow through a narrow two-dimensional channel and thus can be probed experimentally. }

\section{I. Kubo formula calculation and observable signatures}

Conductivity in a clean noninteracting system is given by the Kubo formula $\sigma_{\alpha \beta}(i\omega_n, \bq) =  -K_{\alpha \beta}(i\omega_n, \bq)/\omega_n$~\cite{Colemanbook}, with the current-current correlation function $K_{\alpha \beta}(i\omega_n, \bq) = \langle \hat j^\alpha(i\omega_n, \bq) \hat j^\beta(-i\omega_n, -\bq)  \rangle$ that equals to

\be  
K_{\alpha \beta}(i\omega, \bq) = - \frac1S \sum_{m,n,\bk} \frac{n_F[\ve_n(\bk)] - n_F[\ve_m(\bk+\bq)]}{\ve_n(\bk) - \ve_m(\bk+\bq) + i \hbar \omega} F_{\alpha \beta}^{nm}(\bk,\bq), \label{SMEq:Kdef}
\ee 
where  
\be  
F_{\alpha \beta}^{nm}(\bk,\bq) \equiv \langle u_n(\bk)|\hat j_{\bk+\frac{\bq}2}^\alpha|u_m(\bk+\bq)\rangle \langle u_m(\bk + \bq)|\hat j_{\bk+\frac{\bq}2}^\beta| u_n(\bk)\rangle, \qquad \hat j_\bk^{\alpha} \equiv \frac{e}{\hbar}\frac{\partial \hat H_0(\bk)}{\partial k_\alpha}.
\ee
Here $\ve_n(\bk)$ define the energy bands of the crystal, $S$ is the total area of the system, and $n_F[\ve]$ is the Fermi-Dirac distribution function. An extra minus sign in Eq.~\eqref{SMEq:Kdef} comes from the fermion loop~\cite{Colemanbook}. At zero temperature, $n_F[\ve]$ becomes simply the Heaviside step function, $n_F[\ve] = \Theta(\ve_F-\ve)$, where $\ve_F$ is the Fermi energy.  We emphasize that the correlator $K_{\alpha \beta}(i\omega, \bq)$ only determines the paramagnetic contribution to the total current and, in principle, the diamagnetic term should also be added. The latter, however, is purely symmetric; hence, it does not contribute to the antisymmetric (Hall) part of conductivity and will be neglected hereafter. The summation over $\bk$ should be understood as $\frac1S \sum_{\bk} \to \int \frac{d \bk}{(2\pi)^2}$.

We focus on a two-level system in two dimensions without band crossings and assume that the Fermi level resides in the valence band for definiteness. Then, we separate the total response function into the sum of the interband and intraband contributions, $K_{\alpha \beta}(i\omega, \bq) = K_{\alpha \beta}^{\text{inter}}(i\omega, \bq)+K_{\alpha \beta}^{\text{intra}}(i\omega, \bq)$:

\iffalse
\begin{align}
K_{\alpha \beta}^{\text{inter}}(i\omega, \bq) &= \sum_{\bk \in \text{occ.}}  \frac{ \langle u_V(\bk)|\hat j_{\bk+\frac{\bq}2}^\alpha|u_C(\bk+\bq)\rangle \langle u_C(\bk + \bq)|\hat j_{\bk+\frac{\bq}2}^\beta| u_V(\bk)\rangle}{\ve_V(\bk) - \ve_C(\bk+\bq) + i \omega} + \left[ \omega \to -\omega, \bq \to -\bq, \{\alpha, \beta\} \leftrightarrow \{\beta, \alpha\} \right], \nonumber \\ K_{\alpha \beta}^{\text{intra}}(i\omega, \bq) &= \sum_{\bk \in \text{occ.}}  \frac{ \langle u_V(\bk)|\hat j_{\bk+\frac{\bq}2}^\alpha|u_V(\bk+\bq)\rangle \langle u_V(\bk + \bq)|\hat j_{\bk+\frac{\bq}2}^\beta| u_V(\bk)\rangle}{\ve_V(\bk) - \ve_V(\bk+\bq) + i \omega} + \left[ \omega \to -\omega, \bq \to -\bq, \{\alpha, \beta\} \leftrightarrow \{\beta, \alpha\} \right], \label{Eq:Kintra}
\end{align}
\fi

\begin{align}
K_{\alpha \beta}^{\text{inter}}(i\omega, \bq) &= - \frac1S\sum_{\bk \in \text{occ.}}  \frac{ F_{\alpha \beta}^{VC}(\bk, \bq)}{\ve_V(\bk) - \ve_C(\bk+\bq) + i \hbar \omega} +\frac{ F_{\beta \alpha}^{VC}(\bk, -\bq)}{\ve_V(\bk) - \ve_C(\bk-\bq) - i \hbar \omega}, \nonumber \\ K_{\alpha \beta}^{\text{intra}}(i\omega, \bq) &= - \frac1S\sum_{\bk \in \text{occ.}}  \frac{F_{\alpha \beta}^{VV}(\bk, \bq)}{\ve_V(\bk) - \ve_V(\bk+\bq) + i \hbar \omega} + \frac{F_{\beta \alpha}^{VV}(\bk, -\bq)}{\ve_V(\bk) - \ve_V(\bk-\bq) - i \hbar \omega}, \label{SMEq:Kintra}
\end{align}
where indices ``$V$'' and ``$C$'' stand for ``valence'' and ``conduction'', respectively, and the summation is over the occupied states only. Below, we calculate these contributions separately, so the total Hall conductivity $\sigma_{AH}(\omega,\bq)\equiv [\sigma_{xy}(\omega,\bq) - \sigma_{yx}(\omega,\bq)]/2$ is given  by the sum  $\sigma_{AH}(\omega,\bq) = \sigma_{AH}^{\text{inter}}(\omega,\bq) + \sigma_{AH}^{\text{intra}}(\omega,\bq)$. The subscript ``$AH$'' stands for ``anomalous Hall''.

\subsection{I.A Geometry of the band structure}

Before continuing the calculation, we introduce the Berry curvature $\Omega_{\alpha \beta}(\bk)$, quantum metric $g_{\alpha \beta}(\bk)$, and the (real) symmetric rank-3 tensor $T_{\alpha \beta \gamma}(\bk)$ of the valence band as

\begin{align}  
\Omega_{\alpha \beta}(\bk) &= \frac{\partial A_\beta(\bk)}{\partial k_\alpha} - \frac{\partial A_\alpha(\bk)}{\partial k_\beta} =i\left(  \left< \frac{ \partial u_V(\bk)}{\partial k_\alpha} \middle| \frac{\partial u_V(\bk)}{\partial k_\beta} \right> -  \left< \frac{ \partial u_V(\bk)}{\partial k_\beta} \middle| \frac{\partial u_V(\bk)}{\partial k_\alpha} \right>    \right) = -2 \text{Im}  \left< \frac{ \partial u_V(\bk)}{\partial k_\alpha} \middle| \frac{\partial u_V(\bk)}{\partial k_\beta} \right>, \nonumber \\ g_{\alpha \beta}(\bk) &= \frac12\left\{  \left< \frac{ \partial u_V(\bk)}{\partial k_\alpha} \middle| \frac{\partial u_V(\bk)}{\partial k_\beta} \right> - \left< \frac{ \partial u_V(\bk)}{\partial k_\alpha} \middle| u_V(\bk) \right>\left< u_V(\bk)\middle|\frac{\partial u_V(\bk)}{\partial k_\beta} \right> + (\alpha \leftrightarrow    \beta)  \right\}, \nonumber \\ 
T_{\alpha \beta \gamma}(\bk)  \equiv  &      - \left( A_\gamma g_{\alpha \beta} + A_\alpha g_{\gamma \beta} + A_\beta g_{\alpha \gamma}\right) - A_\alpha A_\beta A_\gamma + \frac13\left( \frac{\partial^2 A_\gamma}{\partial k_\beta \partial k_\alpha} + \frac{\partial^2 A_\alpha}{\partial k_\beta \partial k_\gamma} + \frac{\partial^2 A_\beta}{\partial k_\gamma \partial k_\alpha} \right)   - \nonumber  \\  &-\frac{i}{2} \left[  2\left\langle    u_V \middle| \frac{\partial^3 u_V}{\partial k_\alpha \partial k_\beta \partial k_\gamma} \right\rangle + \frac{\partial(g_{\alpha\beta} + A_\alpha A_\beta)}{\partial k_\gamma} + \frac{\partial(g_{\alpha\gamma} + A_\alpha A_\gamma)}{\partial k_\beta}     + \frac{\partial(g_{\beta\gamma} + A_\beta A_\gamma)}{\partial k_\alpha}\right], \label{SMEq:OmegagT}
\end{align}
where we have also defined the Berry connection $A_\alpha(\bk)\equiv i \left< u_V(\bk) \middle| \frac{\partial u_V(\bk)}{\partial k_\alpha} \right>$ (again, we suppress argument $\bk$ in the expression for $T_{\alpha \beta \gamma}$ for brevity). All the quantities are calculated in the valence band since we assumed that the Fermi energy lies within the valence band. We point out that $\Omega_{\alpha \beta}$, $g_{\alpha \beta}$, and $T_{\alpha \beta \gamma}$ are invariant under the gauge transformations $|u(\bk)\rangle \to e^{i\phi_\bk}|u(\bk) \rangle$. To emphasize this fact, we note that these quantities may be rewritten through other gauge-invariant objects, quantum geometric tensor $c_{\alpha \beta}$ and the  quantum  geometric  connection $c_{\alpha \beta \gamma}$:

\be  \label{SMEq:ccdef}
c_{\alpha \beta} = \langle u_V|(\partial_{k_\alpha} P_C)(\partial_{k_\beta} P_C)|u_V  \rangle, \qquad c_{\alpha \beta \gamma} = \langle u_V|(\partial_{k_\alpha} \partial_{k_\beta}P_C)(\partial_{k_\gamma} P_C)|u_V  \rangle,
\ee
where $P_C = |u_C\rangle \langle u_C| = 1 - |u_V\rangle \langle u_V|$ is the projector onto the conduction band. Then, it is straightforward to check that 

\begin{align}
&c_{\alpha \beta} = \left(g_{\alpha \beta} - \frac{i}2 \Omega_{\alpha \beta} \right), \qquad c_{\alpha \beta \gamma} = - i(A_\alpha g_{\beta \gamma} + A_\beta g_{\alpha \gamma} + A_\gamma g_{\alpha \beta}) - \frac12\left( A_\alpha \Omega_{\beta \gamma} + A_\beta  \Omega_{\alpha \gamma} - A_\gamma \Omega_{\alpha \beta}  \right) + i\frac{\partial^2 A_\gamma}{\partial k_\alpha \partial k_\beta} + \nonumber   \\ &+ \left< u_V \middle| \frac{\partial^3 u_V}{\partial k_\alpha \partial k_\beta \partial k_\gamma}   \right> + \frac{\partial}{\partial k_\beta} \left( g_{\alpha \gamma} + A_\alpha A_\gamma - \frac{i}2 \Omega_{\alpha \gamma} \right) + \frac{\partial}{\partial k_\alpha} \left( g_{\beta \gamma} + A_\beta A_\gamma - \frac{i}2 \Omega_{\beta \gamma} \right) - i A_\alpha A_\beta A_\gamma  - A_\gamma \frac{\partial A_\beta}{\partial k_\alpha}, \label{SMEq:ccfull}
\end{align}
so one easily finds 

\be  
\Omega_{\alpha \beta} = -2\,  \text{Im} \, c_{\alpha \beta}, \qquad g_{\alpha \beta} = \text{Re} \, c_{\alpha \beta}, \qquad T_{\alpha \beta \gamma} =  \frac13 \, \text{Im} \, (c_{\alpha \beta \gamma} + c_{\beta \gamma \alpha} + c_{\gamma \alpha \beta}), \label{SMEq:OmegagTshort}
\ee 
where in the last equality we also used 

\be  
\frac13 (c_{\alpha \beta \gamma} + c_{\beta \gamma \alpha} + c_{\gamma \alpha \beta}) =  i T_{\alpha \beta \gamma} + \frac16\left(  \frac{\partial g_{\beta \gamma}}{\partial k_\alpha} + \frac{\partial g_{\alpha \gamma}}{\partial k_\beta} + \frac{\partial g_{\alpha \beta}}{\partial k_\gamma}\right).
\ee

The real part of $c_{\alpha \beta \gamma}$ is related to the quantum metric as 

\bea
\text{Re} \, c_{\alpha \beta \gamma} = \frac{1}{2} \left( \frac{\partial g_{\beta \gamma} }{\partial k_{\alpha}} + \frac{\partial g_{\alpha \gamma}}{\partial k_{\beta}}  - \frac{\partial g_{\alpha \beta}}{\partial k_{\gamma}}  \right) \equiv g_{\gamma \delta} \Gamma^{\delta}_{\alpha \beta},
\eea
where $\Gamma^{\delta}_{\alpha \beta}$ are the  Christoffel symbols of the Levi-Civita connection of the metric tensor $g_{\alpha \beta}$. Similarly, one can check that 
\bea
\text{Im} \, c_{\alpha \beta \gamma} = \frac{1}{2} \Omega_{\gamma \delta} \Gamma^{\delta}_{\alpha \beta}.
\eea
Additionally, $c_{\alpha \beta \gamma}$ satisfies  the identity $ \partial_{k_\gamma} \Omega_{\alpha \beta} = 2\,\text{Im} \left[ c_{\gamma  \beta \alpha} -c_{\gamma  \alpha \beta}  \right]$  which can be rewritten as 

\bea \label{cov_div}
\nabla_{\gamma} \Omega_{\alpha \beta} = \frac{\partial \Omega_{\alpha \beta}}{\partial k_\gamma} - \Gamma^{\delta}_{\alpha \gamma} \Omega_{\delta \beta} - \Gamma^{\delta}_{\beta \gamma} \Omega_{\alpha \delta} = 0, 
\eea
$\nabla_{\gamma}$ is a covariant derivative. Geometrically, Eq.~(\ref{cov_div}) shows that the Levi-Civita connection (a certain way of parallel transporting vectors) defined via $g_{\alpha \beta}$ preserves the Berry curvature two-form.

To elucidate the geometric origin and properties of the tensor $T_{\alpha \beta \gamma}$ it is useful to introduce the symplectic connection following 
Ref.~\cite{Fedosov:1996fu} (our definition is different from that in Ref.~\cite{AhnNagaosa2020}): $\tilde{\Gamma}_{\gamma \alpha \beta} = \Omega_{\gamma \delta}\tilde{\Gamma}^{\delta}_{\alpha \beta } = \frac{1}{3}(\partial_\beta \Omega_{\gamma \alpha} + \partial_\alpha \Omega_{\gamma \beta})$. By construction, this connection preserves the Berry curvature 2-form and has a vanishing fully-symmetric part. In these terms, $T_{\alpha \beta \gamma}$ is simply given by the difference of the two connections

\bea
T_{\alpha \beta \gamma} = \tilde{\Gamma}_{\gamma \alpha \beta } - \Omega_{\gamma \delta}\Gamma^{\delta}_{\alpha \beta}.
\eea

This representation shows that $T_{\alpha \beta \gamma}$, as any difference of two connections, is a tensor under diffeomorphisms of the Brillouin zone, similar to the rank-2 tensors $g_{\alpha \beta}$ and $\Omega_{\alpha \beta}$. As we will show below in Section~II, it determines the gauge-invariant part of the third cumulant of the position operator, analogous to how the quantum metric $g_{\alpha \beta}$ determines the gauge-invariant part of the second cumulant~\cite{MarzariVanderbilt97,PrunedaSouza09,SouzaWilkensMartin2000,Sgiarovelloetal2001}.

\iffalse
As was shown in Ref.~\cite{Fedosov:1996fu}, any symplectic connection can be represented in the form $\tilde{\Gamma}_{\alpha \beta \gamma} = \frac{1}{3}(\partial_\beta \Omega_{\gamma \alpha} + \partial_\alpha \Omega_{\gamma \beta}) + t_{\alpha \beta \gamma}$, where $t_{\alpha \beta \gamma}$  can be any fully symmetric rank-3 tensor. In our case, $\tilde{\Gamma}_{\alpha \beta \gamma}$ and $t_{\alpha \beta \gamma}$ are fixed  by the band structure  such that this relation takes form 
\be 
\text{Im}\, c_{\alpha \beta \gamma} = -\frac{1}{6}(\partial_\alpha \Omega_{\beta \gamma} + \partial_\beta \Omega_{\alpha \gamma}) + T_{\alpha \beta \gamma}.
\ee
\fi
\iffalse
More specifically, it is equal to the third gauge-invariant cumulant of electron configuration introduced in \cite{SouzaWilkensMartin2000}.

Let us introduce the generating function for gauge-invariant cumulatns

\bea
\log C(\alpha) = \frac{V}{(2 \pi)^3} \int dk \log \langle u_k | u_{k + \alpha} \rangle,
\eea
where V is the volume of the system. The cumulants are defined by differentiating the logarithm of the generating function with respect to $\alpha$
\bea
\langle\hat{X}_c^i \dots \hat{X}_c^j \rangle = i \partial_{\alpha_i} \cdots \partial_{\alpha_j} \log C(\alpha)|_{\alpha = 0}.
\eea
For the first few cumulants we obtain:

\bea
\langle \hat{X}^i \rangle_c = \frac{V}{(2 \pi)^2}  \int_{BZ} dk ~ A^i(k),~ \langle \hat{X}^i \hat{X}^j \rangle_c = \frac{V}{(2 \pi)^2}  \int_{BZ} dk ~ g^{i j}(k),~ \langle \hat{X}^i \hat{X}^j \hat{X}^k \rangle_c = -\frac{V}{(2 \pi)^2} \int_{BZ} dk ~ T^{i j k} (k).
\eea
\fi

\subsection{I.B Interband contribution}

The most involved part of the calculation is the evaluation of the matrix elements $F_{\alpha \beta}^{VC}(\bk, \bq)$ and $F_{\alpha \beta}^{VV}(\bk, \bq)$. In this subsection, we focus on $F_{\alpha \beta}^{VC}(\bk, \bq)$ (interband contribution). Without loss of generality, we choose vector $\bq$ to be along the $x$-axis, $\bq = q \hat x$. Then, one can expand $F_{\alpha \beta}^{VC}(\bk, \bq)$ into the Taylor series: 

\be 
F_{\alpha \beta}^{VC}(\bk, \bq) = \frac{e^2}{\hbar^2}  \left[  f^{VC}_{0\alpha \beta}(\bk) + f^{VC}_{1\alpha \beta}(\bk) q + f^{VC}_{2\alpha \beta}(\bk)  q^2 + \ldots\right].
\ee

The calculation of $f^{VC}_{i\alpha \beta}(\bk)$  is tedious but straightforward, and we find:

\begin{align} 
f^{VC}_{0\alpha \beta}(\bk) &= (\ve_C - \ve_V)^2 c_{\alpha \beta}, \nonumber \\
f^{VC}_{1\alpha \beta}(\bk) &= \left[\frac{(\ve_C - \ve_V)^2}2 c_{\alpha \beta}\right]' +  \frac{(\ve_C - \ve_V)}2 \hbar \left[ (v_{C\alpha} + v_{V \alpha}) c_{x \beta} +   (v_{C\beta} + v_{V \beta})   c_{\alpha x}  \right], \nonumber \\    f^{VC}_{2\alpha \beta}(\bk) &= \left[\frac{(\ve_C - \ve_V)^2}8 c_{\alpha \beta}\right]'' + \frac{(\ve_C - \ve_V)^2}4 \left\{ c_{x \alpha} c_{x \beta} + c_{\alpha x} c_{\beta x} - 2 c_{x x} c_{\alpha \beta} \right\} +  \nonumber \\ &+ \frac{\ve_C - \ve_V}4 \hbar \left\{ \left[ (v_{C\alpha} + v_{V \alpha}) c_{x \beta} + (v_{C\beta} + v_{V \beta}) c_{\alpha x} \right]'  - \frac{v_{C\alpha} - v_{V\alpha}}2 c_{xx \beta} - \frac{v_{C\beta} - v_{V\beta}}2  \bar{c}_{x x \alpha} \right\} +
\nonumber \\ & +\frac{\hbar^2}4 \left\{ (v_{C\alpha}  + v_{V\alpha})(v_{C\beta}  + v_{V\beta}) c_{xx} + (v_{Cx} - v_{Vx})(v_{C\alpha} + v_{V\alpha}) c_{x \beta}  +  (v_{Cx} - v_{Vx})(v_{C\beta} + v_{V\beta}) c_{\alpha x}        \right\}, \label{SMEq:fVC}
\end{align}
where we suppressed argument $\bk$ on the right hand side of all the equalities for brevity, and tensors $c_{\alpha\beta}$ and $c_{\alpha \beta \gamma}$ are defined in Eqs.~\eqref{SMEq:ccdef}-\eqref{SMEq:ccfull}. Symbol ``$\, '\,$'' implies the derivative with the respect to $k_x$, i.e., $f'(\bk) \equiv \partial f(\bk)/\partial k_x$, and we have defined the band velocities $v_{V(C)\alpha} \equiv \partial_{k_\alpha}\ve_{V(C)}/\hbar$.  When deriving the above expressions, we found it useful to exploit the relation 

\be  
\left<u_n(\bk)\middle| \frac{\partial \hat H_0(\bk)}{\partial k_\alpha} \middle| u_m(\bk)\right> = \delta_{nm} \hbar v_{n \alpha}(\bk) + \left[ \ve_m(\bk) - \ve_n(\bk)  \right] \left< u_{n}(\bk) \middle| \frac{\partial u_m(\bk)}{\partial k_\alpha}   \right>, \qquad \left< u_n(\bk) \middle| u_m(\bk) \right> = \delta_{nm},
\ee
where $n$ and $m$ label bands.

Since we are interested in the Hall conductivity, we only need the antisymmetric parts of functions $f_{i\alpha \beta}^{VC}(\bk)$. After antisymmetrization, we find for functions $\tilde f_{i\alpha\beta}^{VC} \equiv f_{i\alpha\beta}^{VC} - f_{i\beta \alpha}^{VC}$:  

\begin{align} \label{SMEq:fVCas}
\tilde f^{VC}_{0\alpha \beta}(\bk) &= -i (\ve_C - \ve_V)^2 \Omega_{\alpha \beta}, \\
\tilde f^{VC}_{1\alpha \beta}(\bk) &= -\frac{i}2 \left[ (\ve_C - \ve_V)^2 \Omega_{\alpha \beta} \right]' + \frac{i}2 \hbar (\ve_C - \ve_V) \left[ (v_{C\beta} + v_{V\beta}) \Omega_{x\alpha} - (v_{C\alpha} + v_{V\alpha}) \Omega_{x\beta} \right], \nonumber \\   \tilde f^{VC}_{2\alpha \beta}(\bk) &= -\frac{i}{16} \left[ (\ve_C - \ve_V)^2 \Omega_{\alpha \beta} \right]'' + \frac{i}4 \hbar \left[ (\ve_C - \ve_V)(v_{C\beta} + v_{V\beta}) \Omega_{x\alpha}\right]' + \frac{i}4(\ve_C - \ve_V)^2 g_{xx} \Omega_{\alpha \beta}+ \nonumber \\ &+ \frac{i}4 \hbar (\ve_C - \ve_V)(v_{C\alpha} - v_{V\alpha})\left( \frac13\Omega_{x\beta}' - T_{xx\beta}  \right) - (\alpha \leftrightarrow \beta). \nonumber
\end{align}

Expanding Eq.~\eqref{SMEq:Kintra} in small $\omega$ and $q$ up to the order $O(\omega q^2, \omega^2 q, \omega^3)$, we find for the antisymmetrized interband contribution to the current-current correlation function,  $\tilde K^{\text{inter}}_{\alpha \beta}(i\omega, q) \equiv [K^{\text{inter}}_{\alpha \beta}(i\omega, q)-K^{\text{inter}}_{\beta \alpha}(i\omega, q)]/2$:

\iffalse
The interband contribution can be evaluated straightforwardly. The answer for the Hall (antisymmetric) part of the conductivity tensor reads as

\be  
K^{\text{inter}}_{\alpha \beta}(i\omega, q) = \omega \sum_{\bk \in \text{occ.}} \left\{ -\Omega_{\alpha \beta}  + \frac{q^2}2 \left[ g_{xx} \Omega_{\alpha \beta} - \frac14 \Omega_{\alpha\beta}^{''} + O\left( \frac1{\ve_C - \ve_V} \right) \right] \right\},
\ee
where $\Omega_{\alpha \beta}$ and $g_{\alpha \beta}$ are the Berry curvature and quantum metrics tensors, respectively, and we assumed that vector $\bq$ is in $x$-direction ($\Omega_{\alpha\beta}^{''}$ implies the second derivative with respect to $k_x$).

After antisymmetrization, $\tilde K^{\text{inter}}_{\alpha \beta}(i\omega, q) \equiv [K^{\text{inter}}_{\alpha \beta}(i\omega, q)-K^{\text{inter}}_{\beta \alpha}(i\omega, q)]/2$, we find to the order $O(\omega^3,  \omega^2 q, \omega q^2, q^3)$:
\fi

\begin{align}  
&\tilde K^{\text{inter}}_{xy}(i\omega, q) \approx -\frac{e^2}{\hbar^2} \frac1S\sum_{\bk \in \text{occ.}} \frac{i}2 q \left[ (\ve_C - \ve_V)\Omega_{xy} \right]' - \hbar \omega\Omega_{xy}  + \frac{\hbar \omega q^2}2\left\{ g_{xx} \Omega_{xy} - \frac{\Omega_{xy}''}4 + \right. \nonumber \\ &\left.
+   \frac{\hbar}{\ve_C - \ve_V}
\left[ (v_{Cx}\Omega_{xy})' - \frac{v_{Cx} - v_{Vx}}3 \Omega_{xy}' - \frac{v_{Cx} - v_{Vx}}2 T_{xxy} + \frac{v_{Cy} - v_{Vy}}2 T_{xxx}  \right] - \hbar^2 \frac{v_{Cx}(v_{Cx} + v_{Vx})}{(\ve_C - \ve_V)^2}\Omega_{xy}\right\} - \nonumber \\ &- i \hbar^2 \omega^2 q \left\{ \frac12 \left( \frac{\Omega_{xy}}{\ve_C - \ve_V} \right)' - \hbar\frac{v_{Cx} + v_{Vx}}{(\ve_C - \ve_V)^2} \Omega_{xy} \right\} + \hbar^3 \omega^3 \frac{\Omega_{xy}}{(\ve_C - \ve_V)^2}.
\end{align}
We note that that we have neglected the $q^3$ contribution here. The interband contribution to the Hall conductivity is determined as $\sigma_{AH}^{\text{inter}}(i\omega_n, \bq)\equiv [\sigma_{xy}^{\text{inter}}(i\omega_n, \bq) - \sigma_{yx}^{\text{inter}}(i\omega_n, \bq)]/2 =  -\tilde K_{xy}^{\text{inter}}(i\omega_n, \bq)/\omega_n$, which after the analytical continuation $i\omega \to \omega+i \delta$ becomes ($\delta$ is an infinitesimal positive number which  physically corresponds to the single-particle scattering rate)

\begin{align}  
&\sigma_{AH}^{\text{inter}}(\omega, q) \approx -\frac{e^2}{\hbar}\frac1S\sum_{\bk \in \text{occ.}} \Omega_{xy} + \frac{q}{2\hbar \omega}  \left[ (\ve_C - \ve_V)\Omega_{xy} \right]'   - \frac{q^2}2\left\{ g_{xx} \Omega_{xy} - \frac{\Omega_{xy}''}4 + \right. \nonumber \\ &\left.
+   \frac{\hbar}{\ve_C - \ve_V}
\left[ (v_{Cx}\Omega_{xy})' - \frac{v_{Cx} - v_{Vx}}3 \Omega_{xy}' - \frac{v_{Cx} - v_{Vx}}2 T_{xxy} + \frac{v_{Cy} - v_{Vy}}2 T_{xxx}  \right] - \hbar^2 \frac{v_{Cx}(v_{Cx} + v_{Vx})}{(\ve_C - \ve_V)^2}\Omega_{xy}\right\} + \nonumber \\ & + \hbar \omega q \left\{ \frac12 \left( \frac{\Omega_{xy}}{\ve_C - \ve_V} \right)' - \hbar \frac{v_{Cx} + v_{Vx}}{(\ve_C - \ve_V)^2} \Omega_{xy} \right\} + \hbar^2 \omega^2 \frac{\Omega_{xy}}{(\ve_C - \ve_V)^2}.
\end{align}

The first term in this expression, $\Omega_{xy}$, is the conventional (uniform) intrinsic contribution to the Hall conductivity. The second term is proportional to $q/\omega$ and, in principle, can be larger than the $q^2$ contribution, which is the main focus of this work. However, this term vanishes in the insulating state after summation over $\bk$ (since it is proportional to a full derivative). Also, as we show below, this term exactly cancels the corresponding $q/\omega$ term from the intraband contribution in the static limit, $\omega\ll v_F q$, where $v_F$ is some typical Fermi velocity at the Fermi level. In the optical limit, $\omega\gg v_F q$, this term must be taken into account in the metallic regime. 

The third term is proportional to $q^2$ and in the insulating regime (when all full derivatives can be discarded) exactly reproduces Eq.~\eqref{Eq:sigma02} of the main text. Finally, the terms proportional to $\omega q$ and $\omega^2$ can be neglected in the static limit or when the band gap, $\ve_C - \ve_V$, is much larger than the bandwidth. Otherwise, these terms must be taken into account as well.

While all the results presented here were derived for the two-band systems only, the general expression for the interband contribution in the case of multiple bands can be found in Ref.~\cite{Zhong_thesis}.

%{\vk (Would be good to understand what happens to other terms in $\tilde K$ of the order $\omega^0$ and higher powers of $q$!)} {\vk Maybe, we should add the expression for the case when Fermi level is in the conduction band.}

\subsection{I.C Intraband contribution}

The calculation of the intraband contribution is similar but much more lengthy. Since there is no band gap $\ve_C - \ve_V$ in the denominator of the intraband term in Eq.~\eqref{SMEq:Kintra}, we need to expand $F_{\alpha \beta}^{VV}(\bk, \bq)$ up to the fourth order in $\bq$. Assuming again that $\bq$ is along the $x$-axis, one can write

\be  
F_{\alpha \beta}^{VV}(\bk, \bq) = \frac{e^2}{\hbar^2} \left[ f^{VV}_{0\alpha \beta}(\bk) + f^{VV}_{1\alpha \beta}(\bk) q + f^{VV}_{2\alpha \beta}(\bk) q^2 + f^{VV}_{3\alpha \beta}(\bk) q^3 + f^{VV}_{4\alpha \beta}(\bk) q^4 + \ldots \right].
\ee
As we are interested in the antisymmetric part only, we directly calculate $\tilde f^{VV}_{i \alpha \beta}(\bk) \equiv f^{VV}_{i \alpha \beta}(\bk) - f^{VV}_{i \beta \alpha}(\bk)$. After lengthy calculation, we find

\begin{align}
&\tilde f^{VV}_{0 \alpha \beta} (\bk) = 0, \qquad \tilde f^{VV}_{1 \alpha \beta}(\bk) = i \hbar [\ve_C(\bk) - \ve_V(\bk)][v_{V\alpha}(\bk) \Omega_{x\beta}(\bk) - v_{V\beta}(\bk) \Omega_{x \alpha}(\bk)], \qquad \tilde f^{VV}_{2 \alpha \beta}(\bk) = \frac12 \left[ \tilde f^{VV}_{1\alpha \beta}(\bk) \right]', \nonumber \\ &\tilde f^{VV}_{3\alpha \beta}(\bk) =\frac{i}8 \hbar \left[ (\ve_C - \ve_V) \Omega_{x\beta} v_{V\alpha} \right]'' - \frac{i}4 \hbar (\ve_C - \ve_V)(v_{V\alpha} + v_{C \alpha}) g_{xx} \Omega_{x\beta} - \frac{i}8 (\ve_C - \ve_V)^2 \Omega_{x\alpha} \left(\frac{\partial g_{xx}}{\partial k_\beta} - 2 g_{x \beta}'   \right) - \nonumber \\ &-\frac{i}4 \hbar^2 v_{V\alpha} v_{C \beta} T_{xxx} + \frac{i(\ve_C - \ve_V)v_{V\alpha}}{4} \hbar \left[ -g_{xx} \Omega_{x\beta} + \frac1{12} \Omega_{x\beta}'' - \frac{T_{xx\beta}'}2  + \frac16 \frac{\partial T_{xxx}}{\partial k_\beta} \right] - (\alpha \leftrightarrow \beta), \nonumber \\ &\tilde f^{VV}_{4\alpha \beta}(\bk) = \frac12 \left[ \tilde f^{VV}_{3\alpha \beta}(\bk)  \right]' - \frac1{24} \left[ \tilde f^{VV}_{1 \alpha \beta}(\bk)   \right]^{(3)}. \label{SMEq:fvv}
\end{align}

Unlike the interband contribution which could be expanded in small $\omega$ and $q$ simultaneously, the intraband term is very sensitive to the order of limits, i.e., whether the system is in the optical or static limit. In the optical limit, $\omega \gg v_F q$, we find to the leading order for $\tilde K^{\text{intra}}_{xy}(\omega, q) \equiv [K^{\text{intra}}_{xy}(\omega, q)-K^{\text{intra}}_{yx}(\omega, q)]/2$

\be  
\tilde K^{\text{intra}}_{xy}(\omega \gg v_F q) = -\frac{e^2}{\hbar^2}  \frac{q^2}{ \hbar \omega} \frac1S\sum_{\bk \in \text{occ.}} \tilde f_{2xy}^{VV}(\bk) = \frac{e^2}{\hbar^2}  \frac{q^2}{2 i \omega} \frac1S\sum_{\bk \in \text{occ.}} \left[(\ve_C - \ve_V) v_{Vx} \Omega_{xy} \right]', 
\ee
which contributes to the Hall conductivity term

\be
\sigma_{\text{Hall}}^{\text{intra}}(\omega \gg v_F q) = - \frac{e^2}{\hbar^2}  \frac{q^2}{2 \omega^2} \frac1S\sum_{\bk \in \text{occ.}} \left[(\ve_C - \ve_V) v_{Vx} \Omega_{xy} \right]'.
\ee
This term is of the order $q^2/\omega^2$ and hence can be neglected compared to the $O(q/\omega)$ interband contribution (though it still can be larger than the $O(q^2)$ interband contribution). 

In the static limit, $\omega\ll v_F q$, the intraband contribution equals (to the linear order in $\omega$ and quadratic order in $q$)

\be  
\tilde K^{\text{intra}}_{xy}(\omega\ll v_F q) \approx i \omega \sigma_{\text{intra}}(q)  + \frac{i}2 \frac{e^2}{\hbar^2}   q \frac1S \sum_{\bk \in \text{occ.}} \left[ (\ve_C - \ve_V) \Omega_{xy}   \right]', 
\ee
with

\begin{align}
&\sigma_{\text{intra}}(q) \approx \sigma_{\text{intra}}^{(0)} + q^2 \sigma_{\text{intra}}^{(2)}, \nonumber \\
&\sigma_{\text{intra}}^{(0)} = -\frac{i}2 \frac{e^2}{\hbar^3} \frac1S\sum_{\bk \in \text{occ.}} \left(  \frac{\tilde f_{1xy}^{VV}}{(v_{Vx}-i \delta)^2} \right)',  \qquad \sigma_{\text{intra}}^{(2)} = -\frac{i}2 \frac{e^2}{\hbar^3} \frac1S\sum_{\bk \in \text{occ.}} \left(  \frac{\tilde f_{3xy}^{VV}}{(v_{Vx}-i \delta)^2} \right)' -\frac{1}{24} \left( \frac{\tilde f^{VV'}_{1xy}}{(v_{Vx}-i\delta)^2} \right)''-  \nonumber \\ &-  \frac5{24}\left( \frac{\tilde f_{1xy}^{VV''}}{(v_{Vx}-i\delta)^2}  \right)' +\frac14 \left[\frac1{v_{Vx}-i\delta} \left( \frac{\tilde f_{1xy}^{VV}}{v_{Vx}-i\delta}  \right)''  \right]' - \frac1{12}\left[\frac1{v_{Vx}-i\delta} \left( \frac{\tilde f_{1xy}^{VV}}{v_{Vx}-i\delta}  \right)'  \right]'',
\end{align}
and $\delta$ originates from the finite scattering rate. We take $\delta \to +0$ since we consider clean noninteracting systems; it is used to regularize integrals over $\bk$. %{\vk (Can we simplify this expression?)}

 For the intraband contribution to the Hall conductivity,  $\sigma_{AH}^{\text{intra}}(\omega, q) \equiv [\sigma_{xy}^{\text{intra}}(\omega, q) - \sigma_{yx}^{\text{intra}}(\omega, q)]/2$, we then find in the static limit

\be
\sigma_{AH}^{\text{intra}}(\omega \ll v_F q) \approx \sigma_{\text{intra}}(q) + \frac{e^2}{2\hbar^2} \frac{q}{\omega} \frac1S\sum_{\bk \in \text{occ.}} \left[ (\ve_C - \ve_V) \Omega_{xy}   \right]'. \label{SMEq:sigmaHallintra}
\ee

The second term exactly cancels the corresponding $q/\omega$ term from the interband contribution.  As expected, the intraband contribution to the Hall conductivity vanishes when the band is fully filled. Formally, it follows from the fact that the intraband term can be written as a full derivative, hence, it is determined by the vicinity of the Fermi surface. 

%{\vk Should we expand the intraband contribution to the higher orders in $\omega$ as well?}

\subsection{I.D Experimental signatures: Hydrodynamic flow through a narrow channel}

Apart from a direct optical measurement, the correction $\sigma_{AH}^{(2)}$ can be probed in the hydrodynamic flow of an electron liquid through a narrow two-dimensional channel.    To show that, we follow a very general derivation from Ref.~\cite{Gromov2017} and assume the applicability of Ohm's law in the static limit, which at a finite wave vector has the form $E_i(\bq) = \rho_{ij}(\bq) j_j(\bq)$. 

We consider the electrons' flow in the $x$-direction along a uniform externally applied electric field $E_x$ in a two-dimensional channel of width $W$ in the $y$-direction. The current component perpendicular to the channel is zero, $j_y = 0$, which implies a nonzero perpendicular electric field $E_y(y)$ due to the anomalous Hall effect in the system. Assuming that the current and the electric field distributions vary slowly on the scale of the lattice constant, one can expand the resistivity tensor up to the second order in small momentum $q$. Then, in the geometry considered here, Ohm's law can be rewritten in the coordinate space as  (implying that $\bq \to -i {\boldsymbol \nabla}$)

\be
    E_x = \left( \rho^{(0)} - \rho^{(2)} \frac{d^2}{d y^2}  \right) j_x(y), \qquad E_y(y) = -  \left( \rho_{AH}^{(0)} - \rho_{AH}^{(2)} \frac{d^2}{d y^2}  \right) j_x(y). \label{SMEq:Ohmslaw}
\ee
Here $\rho^{(0)}$ and $\rho^{(2)}$ correspond to the $xx$ part of the resistivity tensor, $\rho(q) \approx \rho^{(0)}+ q^2 \rho^{(2)}$, while $\rho_{AH}^{(0)}$ and $\rho_{AH}^{(2)}$ describe the anomalous Hall component according to $\rho_{AH}(q) \approx \rho_{AH}^{(0)} + q^2 \rho_{AH}^{(2)}$ (here we assume that vector $\bq$ is along the $y$-direction). For simplicity, we neglect the difference between the antisymmetric Hall part and the actual off-diagonal $xy$ component of the resistivity tensor, as well as the $q$-linear term, leaving a careful analysis of these subtle details to the future study.

In the clean hydrodynamic regime, where the momentum relaxation occurs only at the channel's boundary, $\rho^{(0)} \approx 0$  vanishes, while $\rho^{(2)} \sim \eta_{xx}$ is finite and proportional to the shear viscosity of the electron fluid. Imposing the no-slip boundary conditions $j_x(\pm W/2) = 0$, one finds for the Hall resistance 

\be  
R_{H}= \frac{V_y}{J_x} = -\rho_{AH}^{(0)} - \frac{12 \rho_{AH}^{(2)}}{W^2} \approx \frac1{\sigma_{AH}^{(0)}} \left( 1 - \frac{12 \sigma_{AH}^{(2)}}{W^2 \sigma_{AH}^{(0)}}  \right), \label{SMEq:RH}
\ee 
where we have defined the Hall voltage $V_y = \int_{-W/2}^{W/2} E_y(y) dy$ and total current $J_x = \int_{-W/2}^{W/2} j_x(y) dy$ and have expressed the Hall resistivity components in the clean limit ($\rho_{xx} \to 0$) through the Hall conductivity according to $\rho_{AH}^{(0)} = -1/\sigma_{AH}^{(0)}$ and $\rho_{AH}^{(2)} = \sigma_{AH}^{(2)}/(\sigma_{AH}^{(0)})^2$. This result shows that $\sigma_{AH}^{(2)}$ determines the $1/W^2$ correction to the Hall resistance in the hydrodynamic regime, analogous to how the Hall viscosity $\eta_{xy}$ does it in the Galilean-invariant systems in a nonquantized external magnetic field~\cite{Scaffidietal2017}. This answer is in full agreement with Ref.~\cite{Gromov2017}.

We emphasize that the $1/W^2$ correction appearing in Eq.~\eqref{SMEq:RH} is the direct consequence of the no-slip boundary conditions. These conditions, however, are inapplicable in the insulating regime when the Fermi energy resides in the gap between the bands. Indeed, in this case the intrinsic Hall resistance is expected to be quantized with no size-dependent corrections (except for the exponentially small ones), which is captured by the no-stress boundary conditions~\cite{Gromov2017}. We expect nevertheless that the $1/W^2$ term may be revealed in the nonquantized regime, i.e., when the system is metallic and the Fermi level lies inside one of the energy bands.

\section{II. Semiclassical approach}

\subsection{II.A Semiclassical equations of motion}

Now we compare our results obtained from the Kubo formula with the predictions of the semiclassical approach. To do that, we first derive the gradient expansion for the semiclassical equations of motion for an electron in a crystal in the presence of a slowly varying electrical field $\bE(\br)$ up to the order $\partial^2_{\br}\bE(\br)$. In our derivation, we closely follow the procedure described in detail by Lapa and Hughes in Ref.~\cite{LapaHughes2019}, making exactly the same assumptions. In particular, we focus on a single band only and neglect the possible contributions of the terms coming from the boundary of the Brillouin zone.

We start with the Hamiltonian for a noninteracting particle in the periodic crystalline field $\hat H_0$ in the presence of an external slowly varying electric field $\bE(\br) = - {\boldsymbol \nabla } \varphi(\br)$, which near $\br=0$ can be expanded into the Taylor series:

\be 
\hat H = \hat H_0 - e \varphi(\hat \br), \qquad \varphi (\br) = -E^\mu r_\mu - \frac12 E^{\mu \nu} r_\mu r_\nu - \frac16 E^{\mu \nu \xi} r_\mu r_\nu r_\xi - \ldots
\ee
Tensors $E^{\mu}$, $E^{\mu \nu}$, and $E^{\mu \nu \xi}$ are fully symmetric and do not depend on $\br$. Hamiltonian $\hat H_0$ is the first-quantized version of $\hat H_0(\bk)$ used in the Kubo formula, rewritten in the coordinate space. The eigenstates of $\hat H_0$ are labeled by quasimomentum $\hbar \bk$ and given by $|\psi_\bk\rangle$:

\be  
\hat H_0 |\psi_\bk\rangle  = \ve_\bk |\psi_\bk\rangle , \qquad  |\psi_\bk\rangle = e^{i \bk \cdot \hat\br} |u_\bk\rangle, \qquad \langle \psi_\bk | \psi_{\bk'} \rangle = \delta(\bk - \bk'), \qquad \langle u_\bk | u_\bk\rangle = 1,
\ee
where $u_\bk(\br) = \langle \br|u_\bk\rangle$ has the periodicity of the crystal in the coordinate space. All the inner products involving $|u_\bk\rangle$ or its derivatives imply the real-space integration over the unit cell multiplied by $(2 \pi)^2/S_c$, where $S_c$ is the volume (area) of the unit cell. %{\vk (Make sure that all the hats are there. Compare to Lapa and Hughes.)}

\iffalse
Next, we introduce Berry connection $A_\mu(\bq)$, Berry curvature $\Omega_{\mu \nu}$, and quantum metric $g_{\mu \nu}$:
\be  
A_\mu (\bq) \equiv i \langle u_\bq | \frac{\partial u_\bq}{\partial q_\mu} \rangle, \qquad \Omega_{\mu \nu} = \frac{\partial A_\nu}{\partial q_\mu} - \frac{\partial A_\mu}{\partial q_\nu}, \qquad g_{\mu \nu} = \frac12\left\{  \langle \frac{ \partial u_\bq}{\partial q_\mu} | \frac{\partial u_\bq}{\partial q_\nu} \rangle - \langle \frac{ \partial u_\bq}{\partial q_\mu} | u \rangle\langle u|\frac{\partial u_\bq}{\partial q_\nu} \rangle + (\mu \leftrightarrow    \nu)  \right\}. 
\ee
\fi

The following matrix elements are useful for the further derivation:

\begin{align}
&\langle \psi_\bk | \hat r_\mu| \psi_{\bk'} \rangle = i \frac{\partial}{\partial k_\mu} \delta(\bk - \bk') + A_\mu(\bk) \delta(\bk - \bk'), \nonumber \\
&\langle \psi_\bk | \hat r_\mu \hat r_\nu| \psi_{\bk'} \rangle = \frac{\partial^2}{\partial k_\mu \partial k'_\nu} \delta(\bk - \bk') + iA_\mu(\bk)\frac{\partial}{\partial k_\nu} \delta(\bk - \bk') + iA_\nu(\bk)\frac{\partial}{\partial k_\mu} \delta(\bk - \bk') - \delta(\bk - \bk') \left\langle    u_\bk \middle| \frac{\partial^2 u_\bk}{\partial k_\mu \partial k_\nu}\right\rangle, \nonumber \\
&\langle \psi_\bk | \hat r_\mu \hat r_\nu  \hat r_\xi|\psi_{\bk'} \rangle = -i  \frac{\partial^3}{\partial k_\mu \partial k_\nu \partial k_\xi} \delta(\bk - \bk') - A_\mu(\bk) \frac{\partial^2 \delta(\bk - \bk')}{\partial k_\nu \partial k_\xi} - A_\nu(\bk) \frac{\partial^2 \delta(\bk - \bk')}{\partial k_\mu \partial k_\xi}  - A_\xi(\bk) \frac{\partial^2 \delta(\bk - \bk')}{\partial k_\nu \partial k_\mu} -    \nonumber \\ &-i \left\langle    u_\bk \middle| \frac{\partial^2 u_\bk}{\partial k_\mu \partial k_\nu}\right\rangle \frac{\partial \delta(\bk - \bk')}{\partial k_\xi} -i \left\langle    u_\bk \middle| \frac{\partial^2 u_\bk}{\partial k_\mu \partial k_\xi}\right\rangle \frac{\partial \delta(\bk - \bk')}{\partial k_\nu}   -i \left\langle    u_\bk \middle| \frac{\partial^2 u_\bk}{\partial k_\xi \partial k_\nu}\right\rangle \frac{\partial \delta(\bk - \bk')}{\partial k_\mu}  - \nonumber \\ &- i \delta(\bk - \bk') \left\langle    u_\bk \middle| \frac{\partial^3 u_\bk}{\partial k_\mu \partial k_\nu \partial k_\xi}\right\rangle,
\end{align}
where, again,  $A_\mu (\bk)\equiv i \left< u(\bk) \middle| \frac{\partial u(\bk)}{\partial k_\mu} \right>$ is the Berry connection.

Now we study the dynamics of a wave packet $|\Psi(t)\rangle$ parameterized by $a(\bk,t)$:
\be  
|\Psi (t) \rangle = \int d\bk' a(\bk', t) |\psi_{\bk'}\rangle. 
\ee
We emphasize again that the wave packet $|\Psi(t)\rangle$ is assumed to be constructed from the states within a single band. The function  $a(\bk, t)$ satisfies the normalization condition $\int d\bk \, |a(\bk,t)|^2 = 1$ and obeys the usual Schr\"odinger equation:

\be  
i \hbar \frac{\partial a(\bk,t)}{\partial t} = \ve_\bk a(\bk,t) - e \int d\bk' \langle  \psi_\bk|\varphi(\hat \br)| \psi_{\bk'}\rangle a(\bk',t).
\ee

Next, we derive the equations of motion for the semiclassical coordinate $R_\mu(t)$ and momentum $K_\mu(t)$ defined as 

\be 
R_\mu(t) \equiv \langle \Psi(t) | \hat r_\mu| \Psi(t)  \rangle, \qquad K_\mu(t) \equiv \langle \Psi(t) | \hat k_\mu| \Psi(t)  \rangle, \label{SMEq:RK}
\ee
where we have introduced the quasimomentum operator  $\hbar \hat k_\mu$ as $\hat k_\mu |\psi_\bk\rangle = k_\mu |\psi_\bk\rangle$. Rewriting function $a(\bk,t)$ as $a(\bk,t) = |a(\bk,t)|e^{-i \gamma(\bk, t)}$, we easily find

\be
R_\mu(t) =   \int d\bk\, \tilde R_\mu(\bk,t)|a(\bk,t)|^2, \qquad K_\mu(t) = \int d\bk\, k_\mu |a(\bk,t)|^2, \label{SMEq:RK1}
\ee
where we have also defined 
\be \label{SMEq:tildeR}
\tilde R_\mu (\bk,t) \equiv \frac{i}2 \left[ \frac1{a(\bk, t)} \cdot \frac{\partial a(\bk,t)}{\partial k_\mu} -  \frac1{a^*(\bk, t)}\cdot \frac{\partial a^*(\bk,t)}{\partial k_\mu} \right]  + A_\mu(\bk) = \frac{\partial \gamma(\bk,t)}{\partial k_\alpha} + A_\mu(\bk).
\ee

The equations of motion then read as 

\begin{align}
\dot{K}_\mu(t) &= -\frac{i e}{\hbar} \int d\bk d\bk' k_\mu \left\{a(\bk,t) a^*(\bk',t)\langle \psi_{\bk'}|\varphi(\hat \br)|\psi_\bk \rangle       -a^*(\bk,t) a(\bk',t)\langle \psi_{\bk}|\varphi(\hat \br)|\psi_{\bk'} \rangle \right\},   \nonumber  \\
\dot{R}_\mu(t) &= \frac1{\hbar} \int d\bk \frac{\partial \ve_\bk}{\partial k_\mu} |a(\bk,t)|^2 + \frac{e}{\hbar} \int d \bk d \bk'    \left\{\frac{\partial a(\bk,t)}{\partial k_\mu} a^*(\bk',t)\langle \psi_{\bk'}|\varphi(\hat \br)|\psi_\bk \rangle       +\frac{\partial a^*(\bk,t)}{\partial k_\mu} a(\bk',t)\langle \psi_{\bk}|\varphi(\hat \br)|\psi_{\bk'} \rangle \right\} - \nonumber \\ &- \frac{i e}{\hbar} \int d\bk d\bk' A_\mu(\bk) \left\{a(\bk,t) a^*(\bk',t)\langle \psi_{\bk'}|\varphi(\hat \br)|\psi_\bk \rangle       - a^*(\bk,t) a(\bk',t)\langle \psi_{\bk}|\varphi(\hat \br)|\psi_{\bk'} \rangle \right\}.
\end{align}
Next, we expand $\varphi(\br)$ in powers of $\br$ and evaluate the above equations order by order in gradients. We write the answer in the form 

\begin{align}
\dot{R}_\alpha(t) &= \dot{R}_\alpha^{(0)}(t) +  E^\mu \dot{R}^{(1)}_{\alpha \mu}(t) + \frac12 E^{\mu\nu} \dot{R}^{(2)}_{\alpha \mu \nu}(t) + \frac16 E^{\mu\nu \xi} \dot{R}^{(3)}_{\alpha \mu \nu \xi}(t)+\ldots \nonumber \\
\dot{K}_\alpha(t) &= E^\mu \dot{K}^{(1)}_{\alpha \mu}(t) + \frac12 E^{\mu\nu} \dot{K}^{(2)}_{\alpha \mu \nu}(t) + \frac16 E^{\mu\nu \xi} \dot{K}^{(3)}_{\alpha \mu \nu \xi}(t)+\ldots
\end{align}

After straightforward calculation, we find 

\begin{align} 
&\dot{K}^{(1)}_{\alpha \mu}(t)= -\frac{e}{\hbar} \delta_{\alpha \mu},     \qquad \dot{K}^{(2)}_{\alpha \mu \nu}(t)= -\frac{e}{\hbar}\left[R_\mu(t) \delta_{\alpha \nu} + R_\nu(t) \delta_{\alpha \mu}  \right],     \nonumber \\
&\dot{K}^{(3)}_{\alpha \mu \nu \xi}(t)= - \frac{e}{\hbar} \left\{\delta_{\alpha \xi} \langle \Psi(t)|\hat r_\mu \hat r_\nu |\Psi(t)   \rangle          +\delta_{\alpha \mu} \langle \Psi(t)|\hat r_\xi \hat r_\nu |\Psi(t)   \rangle      + \delta_{\alpha \nu} \langle \Psi(t)|\hat r_\mu  \hat r_\xi |\Psi(t)   \rangle  \right\},
\end{align}
where $R_\mu(t)$ is given by Eqs.~\eqref{SMEq:RK1}-\eqref{SMEq:tildeR}, and 
\be  
\langle \Psi(t)|\hat r_\mu \hat r_\nu |\Psi(t)   \rangle = \int d\bk \left[ \tilde R_\mu(\bk, t) \tilde R_\nu(\bk,t) + g_{\mu \nu}(\bk) \right]|a(\bk,t)|^2  + \frac1{4|a(\bk,t)|^2} \frac{\partial |a(\bk,t)|^2}{\partial k_\mu} \frac{\partial |a(\bk,t)|^2}{\partial k_\nu}.   \label{SMEq:<rr>}
\ee

The calculation for $\dot R_\alpha(t)$ is also straightforward, but much more tedious. After some work, we find

\begin{align}
\dot R^{(0)}_\alpha(t) &= \frac1{\hbar} \int d\bk \, \frac{\partial \ve_\bk}{\partial k_\alpha} |a(\bk,t)|^2, \qquad \dot R^{(1)}_{\alpha \mu}(t) = -\frac{e}{\hbar} \int d\bk \, \Omega_{\mu \alpha}(\bk) |a(\bk,t)|^2,    \nonumber \\  \dot R^{(2)}_{\alpha \mu \nu}(t) &= -\frac{e}{\hbar}\int d\bk \, \left\{   \tilde R_\nu(\bk,t) \Omega_{\mu \alpha}(\bk) + \tilde R_\mu(\bk,t) \Omega_{\nu \alpha}(\bk)  - \frac{\partial g_{\mu \nu}(\bk)}{\partial k_\alpha}  \right\}|a(\bk,t)|^2,   \nonumber \\ \dot R^{(3)}_{\alpha \mu \nu \xi}(t)&= -\frac{e}{\hbar} \int d\bk \, \left\{ \tilde R_\nu \tilde R_\mu \Omega_{\xi \alpha} + \tilde R_\nu \tilde R_\xi \Omega_{\mu \alpha}  + \tilde R_\xi \tilde R_\mu \Omega_{\nu \alpha}  - \tilde R_\xi \frac{\partial g_{\mu\nu}}{\partial k_\alpha} - \tilde R_\mu \frac{\partial g_{\xi\nu}}{\partial k_\alpha} -\tilde R_\nu \frac{\partial g_{\mu\xi}}{\partial k_\alpha}  +   T_{\alpha \mu \nu \xi} \right\}|a|^2 +    \nonumber \\ &+\frac1{4|a|^2} \left\{ \frac{\partial |a|^2}{\partial k_\mu} \frac{\partial |a|^2}{\partial k_\nu} \Omega_{\xi\alpha}    + \frac{\partial |a|^2}{\partial k_\mu} \frac{\partial |a|^2}{\partial k_\xi} \Omega_{\nu\alpha}  + \frac{\partial |a|^2}{\partial k_\nu} \frac{\partial |a|^2}{\partial k_\xi} \Omega_{\mu\alpha}  \right\}, \label{SMEq:Rdot}
\end{align}
and we have suppressed the indices $\bk$ and $t$ in the expression for  $\dot R^{(3)}_{\alpha \mu \nu \xi}(t)$ for brevity. Tensor $T_{\alpha \mu \nu \xi}$ is defined as

\be 
T_{\alpha \mu \nu \xi} = -\frac{\partial   T_{\mu \nu \xi}}{\partial k_\alpha}   + g_{\nu \mu}\Omega_{\xi \alpha} + g_{\nu \xi} \Omega_{\mu \alpha} + g_{\xi \mu} \Omega_{\nu \alpha}- \frac13\left( \frac{\partial^2 \Omega_{\xi \alpha }}{\partial k_\nu \partial k_\mu} + \frac{\partial^2 \Omega_{\mu \alpha}}{\partial k_\nu \partial k_\xi} + \frac{\partial^2 \Omega_{\nu \alpha}}{\partial k_\xi \partial k_\mu}  \right),  
\ee 
and tensor $T_{\mu \nu \xi}(\bk)$ is defined by Eqs.~\eqref{SMEq:OmegagT}-\eqref{SMEq:OmegagTshort}, as well as the Berry curvature $\Omega_{\mu \nu}(\bk)$ and quantum metric $g_{\mu \nu}(\bk)$.

The above equations of motion are exact in a sense that they describe the evolution of the correlation functions for any shape of the wave packet $a(\bk,t)$ (under the assumption that the wave packet is entirely composed of the states within the same band). They allow for the simplest physical interpretation in the limit when the wave packet is sharply peaked in momentum space, i.e., represents a particle with a well-defined momentum and is given by $|a(\bk,t)|^2 \approx \delta(\bk - \bK)$. The electrical current in this case is simply given by the sum over all occupied states $j_\alpha = -(e/S) \sum_{\bK \in \text{occ}} \dot R_\alpha(\bK) f_\bK$, where $f_\bK$ is the Fermi-Dirac distribution function, and we exactly reproduce Eq.~\eqref{Eq:EOM} of the main text. In particular, terms that explicitly contain $\bR \approx \tilde \bR(\bK)$ represent the Taylor expansion for $E^\mu(\bR)$ and $\partial E^\mu(\bR)/\partial R_\nu$ near $\bR =0$, i.e., $E^\mu(\bR) \approx E^\mu + E^{\mu \nu} R_{\nu} + \frac12 E^{\mu \nu \xi}R_{\nu} R_{\xi}$ and $\partial E^\mu(\bR)/\partial R_\nu \approx  E^{\mu \nu}  + E^{\mu \nu \xi} R_{\xi}$.

We see that the semiclassical approach works best in the insulating regime in the case when energy bands are well-separated. Indeed, in this case, the $q^2$ component of the Hall conductivity is primarily determined by the term $g_{xx} \Omega_{xy}$, see Eq.~\eqref{Eq:sigma02}, which is correctly captured by the semiclassical expressions~\eqref{Eq:EOM} or~\eqref{SMEq:Rdot}. This is not surprising since traditionally semiclassics is designed for a single-band description, hence, not suitable for capturing the terms that explicitly contain the energy gap $\ve_C - \ve_V$. As was demonstrated in Refs.~\cite{GaoYangNiu14,GaoXiao2019}, the terms with the inverse powers of the band gap, like those in Eq.~\eqref{Eq:sigma02} apart from $g_{xx} \Omega_{xy}$, can in principle be captured by semiclassics as the perturbative corrections. We do not perform such analysis in the present work, however.

The agreement of the semiclassical approach with the Kubo formula is much less accurate in the metallic regime. While semiclassics captures certain terms which have the form of full derivatives and thus are absent in the insulating case, it generally poorly reproduces the Kubo formula result. The main reason for that is the presence of a nonzero intraband contribution. As is clear from Eqs.~\eqref{SMEq:fvv}-\eqref{SMEq:sigmaHallintra}, the intraband contribution in the static limit contains terms proportional to the band gap $\ve_C - \ve_V$ or even $(\ve_C - \ve_V)^2$, which clearly could not be captured by semiclassics. 

Finally, the semiclassical equations of motion in the nonuniform electric field suffer from the terms originating from the Heisenberg uncertainty principle, ($\partial_{k_\mu}|a|)(\partial_{k_\nu}|a|)$ and $(\partial_{k_\mu}|a|)(\partial_{k_\nu}|a|) \Omega_{\xi \alpha}$. These terms vanish when dealing with the localized Wannier functions, but become divergent in the case of the wave packets narrow in momentum space, which correspond to the particles with well-defined momenta. While these terms have clear physical meaning when considering the time evolution of the correlation functions, it is not clear how to relate them to the physical observables, such as electrical current. We see, however, that once these wave-packet dependent terms are discarded, the semiclassical equations well agree with the microscopic Kubo formulation in the limit where semiclassics is expected to work, i.e., in the insulating regime with the large band gap.

\subsection{II.B Intuitive interpretation of the semiclassical result}

Now we demonstrate that the semiclassical equations of motion can be obtained from a physically transparent argument. First, we notice that the equation for $\dot \bK$ represents the Newton's second law and can be rewritten as 

\be  
\dot K_\alpha =  -\frac{e}\hbar \langle \Psi(t)| \bE(\hat \br)|\Psi(t)\rangle.
\ee

Second, to derive the equation for $\dot \bR$, we introduce the effecitve total energy of the wave packet as~\cite{LapaHughes2019}

\be 
\varepsilon_{eff} = \langle \Psi(t)| \hat H_0 - e \varphi(\hat \br)|\Psi(t)\rangle = \langle \Psi(t)|\hat H_0| \Psi(t) \rangle + e\langle \Psi(t)|E^\mu \hat r_\mu + \frac12 E^{\mu \nu} \hat r_\mu \hat r_\nu + \frac16 E^{\mu \nu \xi}\hat  r_\mu \hat  r_\nu \hat r_\xi + \ldots |\Psi(t)\rangle.
\ee
To evaluate this expression up to the second order in the electric field gradients, we use matrix elements given by Eqs.~\eqref{SMEq:RK}-\eqref{SMEq:RK1} and~\eqref{SMEq:<rr>} as well as the expression for the third moment

\begin{align}
\langle \Psi(t)|\hat r_\mu \hat r_\nu \hat r_\xi|\Psi(t)   \rangle &= \int d\bk  \left\{  \tilde R_\mu \tilde R_\nu \tilde R_\xi  +\tilde R_\mu g_{\nu \xi} + \tilde R_\nu g_{\xi \mu} + \tilde R_\xi g_{\mu \nu}   -\frac13\left( \frac{\partial^2 \tilde R_\mu}{\partial k_\nu \partial k_\xi}    + \frac{\partial^2 \tilde R_\nu}{\partial k_\xi \partial k_\mu}    + \frac{\partial^2 \tilde R_\xi}{\partial k_\mu \partial k_\nu}  \right) + T_{\mu\nu\xi}\right\} |a|^2 + \nonumber \\      & +  \frac1{4|a|^2} \frac{\partial |a|^2}{\partial k_\mu} \frac{\partial |a|^2}{\partial k_\nu}\tilde R_\xi  + \frac1{4|a|^2} \frac{\partial |a|^2}{\partial k_\nu} \frac{\partial |a|^2}{\partial k_\xi}\tilde R_\mu + \frac1{4|a|^2} \frac{\partial |a|^2}{\partial k_\xi} \frac{\partial |a|^2}{\partial k_\mu}\tilde R_\nu,
\end{align}
where we suppressed index $\bk$ for brevity and $\tilde \bR$ is given by Eq.~\eqref{SMEq:tildeR}.

Focusing again on the wave packets describing particles with the well-defined momenta, $|a(\bk)|^2 \approx \delta(\bk - \bK)$, we find

\begin{align}
&\varepsilon_{eff}(\bR, \bK) = \ve_\bK + e E^{\mu} R_{\mu} + \frac{e}2 E^{\mu \nu} \left( R_{\mu} R_{\nu} + g_{\mu \nu}(\bK) + \int d\bk \frac{\partial |a(\bk)|}{\partial k_{\mu}} \cdot  \frac{\partial |a(\bk)|}{\partial k_{\nu}} \right) + \nonumber \\ &+ \frac{e}6 E^{\mu \nu \xi}  \left[ R_{\mu} R_{\nu} R_{\xi} + R_{\mu} g_{\nu \xi}(\bK) + R_{\nu} g_{\mu \xi}(\bK) + R_{\xi} g_{\nu \mu}(\bK) -\frac13\left( \frac{\partial^2 R_\mu}{\partial K_\nu \partial K_\xi}    + \frac{\partial^2  R_\nu}{\partial K_\xi \partial K_\mu}    + \frac{\partial^2  R_\xi}{\partial K_\mu \partial K_\nu}  \right) + T_{\mu \nu \xi}(\bK) +  \right. \nonumber \\ &\left. \int d\bk \left(\frac{\partial |a(\bk)|}{\partial k_\mu} \cdot \frac{\partial |a(\bk)|}{\partial k_\nu}\tilde R_\xi(\bk)  + \frac{\partial |a(\bk)|}{\partial k_\nu} \cdot  \frac{\partial |a(\bk)|}{\partial k_\xi}\tilde R_\mu(\bk) + \frac{\partial |a(\bk)|}{\partial k_\xi} \cdot \frac{\partial |a(\bk)|}{\partial k_\mu}\tilde R_\nu(\bk) \right) \right],
\end{align}
where we have used $\langle \Psi(t)|\hat H_0| \Psi(t) \rangle = \int d\bk\, \ve_\bk \, |a(\bk)|^2 \approx \ve_\bK $ and $\bR = \langle \Psi(t)| \hat \br|\Psi(t)\rangle \approx \tilde \bR(\bK)$. Terms with $E^{\mu}R_{\mu}$, $E^{\mu\nu}R_{\mu}R_\nu$, and $E^{\mu\nu\xi}R_{\mu}R_\nu R_\xi$ simply sum up into $-\varphi(\bR)$. Treating then $\bR$ and $\bK$ as independent variables,  we reproduce the Newton's second law as

\be  
\frac{\partial \ve_{eff}(\bR, \bK)}{\partial R_\alpha} \approx e E_\alpha(\bR) + \frac{e}2 E_{\alpha \mu \nu} \left(g^{\mu \nu}(\bK) + \int d\bk \frac{\partial |a(\bk)|}{\partial k_{\mu}} \cdot  \frac{\partial |a(\bk)|}{\partial k_{\nu}} \right) = -\hbar \dot K_\alpha,
\ee 
where in order to obtain the last term we used the equality

\be  
\frac{\partial}{\partial R_\alpha}\int d\bk \frac{\partial |a(\bk)|}{\partial k_\mu} \cdot \frac{\partial |a(\bk)|}{\partial k_\nu}\tilde R_\xi(\bk) \approx \delta_{\alpha \xi} \int d\bk \frac{\partial |a(\bk)|}{\partial k_\mu} \cdot \frac{\partial |a(\bk)|}{\partial k_\nu}.
\ee

Analogously, we find

\be  
\frac{\partial \ve_{eff}}{\partial K_\alpha} \approx \frac{\partial \ve_\bK}{\partial K_\alpha} + \frac{e}2 \frac{\partial E^{\mu}(\bR)}{\partial R^{\nu}} \cdot \frac{\partial g_{\mu \nu}(\bK)}{\partial K_\alpha} + \frac{e}6 E^{\mu \nu \xi}\frac{\partial T_{\mu\nu\xi}(\bK)}{\partial K_\alpha},
\ee 
leading to 

\begin{align}
 &\frac{\partial \ve_{eff}(\bR, \bK)}{\partial K_\alpha} - \hbar \Omega_{\alpha \xi} \dot K_{\xi} \approx \frac{\partial \ve_\bK}{\partial K_\alpha} + \frac{e}2 \frac{\partial E^{\mu}(\bR)}{\partial R^{\nu}} \cdot \frac{\partial g_{\mu \nu}(\bK)}{\partial K_\alpha} + \frac{e}6 E^{\mu \nu \xi}\frac{\partial T_{\mu\nu\xi}(\bK)}{\partial K_\alpha} + \nonumber \\ &+ \Omega_{\alpha \xi}(\bK)\left[e E_\xi(\bR) + \frac{e}2 E_{\xi \mu \nu} \left(g^{\mu \nu}(\bK) + \int d\bk \frac{\partial |a(\bk)|}{\partial k_{\mu}} \cdot  \frac{\partial |a(\bk)|}{\partial k_{\nu}} \right) \right] = \nonumber \\ &=\hbar \dot R_\alpha + \frac{e}2 E^{\mu \nu \xi}\left\{ \frac13 \frac{\partial^2 \Omega_{\alpha \xi}(\bK)}{\partial K_\mu \partial K_\nu} + \int d\bk \frac{\partial |a(\bk)|}{\partial k_{\mu}} \cdot  \frac{\partial |a(\bk)|}{\partial k_{\nu}} [\Omega_{\alpha \xi}(\bK)  - \Omega_{\alpha \xi}(\bk)] \right\}.
\end{align}

It is straightforward to show that the expression in the brackets of the last line equals to a certain wave packet-dependent combination of the second derivatives of the Berry curvature and, generally, is nonzero. In fact, the first term can  be entirely absorbed by the second one after the proper redefinition of $|a(\bk)|$. This implies that this term vanishes upon integration over the entire Brillouin zone. At the same time, we recall that semiclassics is expected to agree with the exact answer only in the case when the band is fully filled, provided the wave packet-dependent terms are neglected. With this notion in mind, the above equation can be somewhat loosely rewritten as

%At the same time, we recall that semiclassics is expected to agree with the exact answer only in the case when the band is fully filled, provided the wavepacket-dependent terms are neglected. Hence, all second derivatives of the Berry curvature vanish after the integration over the entire Brillouin zone. With this notion in mind, the above equation can be somewhat loosely rewritten as

\be  
\hbar \dot R_\alpha  \simeq \frac{\partial \ve_{eff}(\bR, \bK)}{\partial K_\alpha} - \hbar \Omega_{\alpha \xi} \dot K_{\xi}.
\ee
We see that the semiclassical equations of motion have exactly same form as in the uniform electric field, provided the effective quasiparticle energy is properly defined.

\iffalse
==========================

It is clear that the term containing derivatives of $|a(\bk)|$ may exactly cancel the second derivative of the Berry curvature $\Omega_{\alpha \xi}$ by the proper choice of the shape of the wave packet. Indeed, choosing the circularly symmetric wave packet of the width $\Delta k$ narrowly peaked at $\bK$, we find

\begin{align}  
&\int d\bk \frac{\partial |a(\bk)|}{\partial k_{\mu}} \cdot  \frac{\partial |a(\bk)|}{\partial k_{\nu}} = \frac{d_0}{(\Delta k)^2}\delta_{\mu \nu}, \qquad \int d\bk \frac{\partial |a(\bk)|}{\partial k_{\mu}} \cdot  \frac{\partial |a(\bk)|}{\partial k_{\nu}}(k-K)_{\eta} = 0, \nonumber \\ &\int d\bk \frac{\partial |a(\bk)|}{\partial k_{\mu}} \cdot  \frac{\partial |a(\bk)|}{\partial k_{\nu}}(k-K)_{\eta} (k-K)_{\zeta} = d_1(\delta_{\mu \eta}\delta_{\nu \zeta} + \delta_{\mu \zeta}\delta_{\nu \eta}) + d_2 \delta_{\mu \nu}\delta_{\eta \zeta}.
\end{align}
All higher moments are proportional to some positive powers of $\Delta k$ and hence can be neglected in the limit $\Delta k \to 0$. Assuming then that $\Omega_{\alpha \xi}(\bk)$ is a slowly varying function of $\bk$ on a scale $\Delta k$ and expanding it in the Taylor series near $\bK$, we obtain

\be   
\int d\bk \frac{\partial |a(\bk)|}{\partial k_{\mu}} \cdot  \frac{\partial |a(\bk)|}{\partial k_{\nu}} [\Omega_{\alpha \xi}(\bK)  - \Omega_{\alpha \xi}(\bk)] \approx -d_1\frac{\partial^2 \Omega_{\alpha \xi}(\bK)}{\partial K_\mu \partial K_\nu} - \frac{d_2}2 \delta_{\mu \nu} \nabla^2 \Omega_{\alpha \xi}(\bK)
\ee 
================================
\fi

Finally, assuming again that the wave packet is narrowly peaked in momentum space, $|a(\bk)|^2 \approx \delta(\bk - \bK)$, we find for the third cumulant

\begin{align}
&\langle \Psi(t)|\delta \hat r_\mu \delta \hat r_\nu \delta \hat r_\xi|\Psi(t)\rangle \approx T_{\mu \nu \xi}(\bK)    -\frac13 \left( \frac{\partial^2 \tilde R_{\xi}(\bK)}{\partial K_\mu \partial K_\nu} + \frac{\partial^2 \tilde R_{\mu}(\bK)}{\partial K_\xi \partial K_\nu} + \frac{\partial^2 \tilde R_{\nu}(\bK)}{\partial K_\mu \partial K_\xi} \right) +  \\ &+\int d\bk \left[ \frac{\partial |a(\bk)|}{\partial k_{\mu}} \cdot  \frac{\partial |a(\bk)|}{\partial k_{\nu}} (\tilde R_\xi  - R_\xi) + \frac{\partial |a(\bk)|}{\partial k_{\xi}} \cdot  \frac{\partial |a(\bk)|}{\partial k_{\nu}} (\tilde R_\mu  - R_\mu)  + \frac{\partial |a(\bk)|}{\partial k_{\mu}} \cdot  \frac{\partial |a(\bk)|}{\partial k_{\xi}} (\tilde R_\nu  - R_\nu) \right], \nonumber 
\end{align}
where $\delta \hat r_\mu \equiv \hat r_\mu - R_\mu = \hat r_\mu - \langle \hat r_\mu \rangle$. While the cumulant generally depends on the second derivatives of $\tilde \bR$, its gauge-invariant part (independent of $\tilde \bR$) is exactly given by $T_{\mu \nu \xi}(\bK)$.

\end{widetext}

\end{document}